\documentclass[a4paper,fleqn,table,xcdraw,expansion]{cas-sc}\PassOptionsToPackage{table,xcdraw}{color}
\usepackage[numbers]{natbib}
\usepackage{xurl} 

\usepackage[expansion]{microtype}

\usepackage{amsmath}
\usepackage{amssymb}
\usepackage{booktabs}

\usepackage{enumitem}
\usepackage{eurosym}
\usepackage{graphicx}
\usepackage[utf8]{inputenc}
\usepackage[english]{babel}
\usepackage[labelfont=bf]{caption}
\usepackage{multicol}
\usepackage{multirow}
\usepackage{pdflscape}
\usepackage{siunitx}
\usepackage{svg}
\usepackage{tabularray}
\usepackage{tikz}
\usetikzlibrary{shapes,backgrounds}
\usepackage{verbatim}
\usepackage{subcaption}
\usepackage{adjustbox}
\usepackage{setspace}
\usepackage[table,xcdraw]{color}

\usepackage{appendix}

\usepackage{nomencl}
\usepackage{mdframed}
\makenomenclature

\setcitestyle{authoryear,open={(},close={)}} 
    
\usepackage[acronym]{glossaries}  
\usepackage{hyperref}  
\usepackage{longtable}  
\makeglossaries

\newacronym{ccs}{CCS}{Carbon Capture and Storage}
\newacronym{co2}{CO$_2$}{Carbon Dioxide}
\newacronym{eu}{EU}{European Union}
\newacronym{euets}{EU ETS}{European Emission Trading System}
\newacronym{h2}{H$_2$}{Hydrogen gas}
\newacronym{ng}{NG}{Natural Gas}
\newacronym{rfnbo}{RFNBO}{Renewable Fuels of Non-Biological Origin}
\newacronym{vres}{VRES}{Variable Renewable Energy Sources}

\def\tsc#1{\csdef{#1}{\textsc{\lowercase{#1}}\xspace}}
\tsc{WGM}
\tsc{QE}

\microtypesetup{}
\begin{document}
\let\WriteBookmarks\relax
\def\floatpagepagefraction{1}
\def\textpagefraction{.001}

\shorttitle{From Policy to Practice: Upper Bound Cost Estimates of Europe’s Green Hydrogen Ambitions}    

\shortauthors{E. Hordvei, S. Hummelen, M. Petersen, S. Backe, P. Granado}  

\title [mode = title]{From Policy to Practice: Upper Bound Cost Estimates of Europe’s Green Hydrogen Ambitions}  

\author[a]{Erlend Hordvei} \credit{Conceptualization, Methodology, Software, Validation, Formal analysis, Writing – original draft preparation, Visualization} 
\author[a]{Sebastian Emil Hummelen} \credit{Conceptualization, Methodology, Software, Validation, Formal analysis, Writing – original draft preparation, Visualization} 
\author[b,d]{Marianne Petersen}[orcid=0009-0000-0698-511X]\corref{cor1} \ead{mariper@dtu.dk} \credit{Conceptualization, Methodology, Validation, Formal analysis, writing - review \& editing, Visualization, Supervision} 
\author[a,c]{Stian Backe} \credit{Conceptualization, Methodology, Software, Validation, Formal analysis, writing - review \& editing, Supervision}
\author[a]{Pedro Crespo del Granado} \credit{supervision, writing - review \& editing}

\cortext[cor1]{Corresponding author}

\affiliation[a]{organization={Dept.~of Industrial Economics and Technology Management, Norwegian University of Science and Technology},
            addressline={Alfred Getz vei 3}, 
            city={Trondheim},
            postcode={7034}, 
            country={Norway}}

\affiliation[b]{organization={Dept.~of Wind and Energy Systems, Technical University of Denmark},
            addressline={Anker Engelunds Vej 1}, 
            city={Kongens Lyngby},
           postcode={2800}, 
            country={Denmark}}

\affiliation[c]{organization={SINTEF Energy Research},
            addressline={Sem Sælands vei 11}, 
            city={Trondheim},
            postcode={7034}, 
            country={Norway}}

\affiliation[d]{organization={Siemens Gamesa Renewable Energy A/S},
            addressline={Borupvej 16}, 
            city={Brande},
            postcode={7330}, 
            country={Denmark}}

\begin{abstract}
As the European countries strive to meet their ambitious climate goals, renewable hydrogen has emerged to aid in decarbonizing energy-intensive sectors and support the overall energy transition. To ensure that hydrogen production aligns with these goals, the European Commission has introduced criteria for additionality, temporal correlation, and geographical correlation. These criteria are designed to ensure that hydrogen production from renewable sources supports the growth of renewable energy. This study assesses the impact of these criteria on green hydrogen production, focusing on production costs and technology impacts. The European energy market is simulated up to 2048 using stochastic programming, applying these requirements exclusively to green hydrogen production without the phased-in compliance period outlined in the EU regulations. The findings show that meeting the criteria will increase expected system costs by €82 billion from 2024 to 2048, largely due to the rapid shift from fossil fuels to renewable energy. The additionality requirement, which mandates the use of new renewable energy installations for electrolysis, proves to be the most expensive, but also the most effective in accelerating renewable energy adoption. 
\end{abstract}

\begin{highlights}
\item Analyzes new EU green hydrogen regulations using the EMPIRE optimization model.
\item Green hydrogen regulations could increase system costs by €82 billion from 2024 to 2048.
\item Most of the cost increases due to new renewable investments in the coming decade.
\item Using existing renewable capacities for green hydrogen can cut costs.
\item More electrolyzers are needed to adapt to renewable availability for green hydrogen.
\end{highlights}


\maketitle

\section{Introduction}\label{ch:intro}
The European Commission has defined green hydrogen, providing clarity to producers and consumers. This definition is outlined in the Delegated Acts on Renewable Hydrogen, supplementing The Revised Renewable Energy Directive (Directive (EU) 2018/2001) \citep{euRenewableHydrogen2023, REDII2018}. It includes requirements for the power supply to electrolysis: additionality (renewable energy from new installations within 36 months of the launch of the hydrogen plant), temporal correlation (production within the same hour as renewable powconsumption)n)and spatial correlation (location within the same power market bidding zone).

To increase green hydrogen uptake, the European Hydrogen Bank launched a pilot auction with an budget of € 800 million \citep{EU2023InnovationFund}. The 132 bids from 17 countries indicate the willingness to scale up production with subsidies \citep{EU2024HydrogenBankAuction}. However, the new definition of green hydrogen in Europe may increase costs, potentially reducing its competitiveness as an energy carrier.

Several multicarrier energy system models with European scope have explicitly modeled hydrogen production from electrolysis. However, the literature review reveals inconsistencies in ensuring the renewable nature of the power supplied to electrolysis. To the authors' knowledge, no quantitative studies exist on the impact of new green hydrogen production rules within the \gls{eu}.

This paper aims to study the impact of the new \gls{eu} definition of green hydrogen production on the European energy system towards 2048, focusing on increased system costs when enforcing all new electrolysis plants to produce green hydrogen. Although "\textit{renewable hydrogen is the most compatible option with the EU’s climate [...] goal}" \cite{european2020hydrogen}, it is not the only \gls{eu} strategy. The \gls{eu} strategy also includes hydrogen production from low-carbon electricity like nuclear power and gas reformation with carbon capture and storage. This study explores the consequences of a renewable hydrogen scenario, providing insights for ongoing political discussions. The new requirements add costs to the future European energy system, necessitating changes in \gls{vres} and electrolysis investments in terms of capacities and geographic locations. Instead of a gradual phase-in, all regulatory criteria are applied immediately, exploring the upper bound system cost estimate and analyzing the hydrogen production system's response to these regulations.

To quantitatively estimate the impact of enforcing green hydrogen production in Europe, the European Model for Investment in Power Systems with high shares of Renewable Energy (EMPIRE) is extended to facilitate electrolysis production that is in large part compatible with the current definition of green hydrogen \gls{eu}. EMPIRE minimizes the total system costs of the European energy system. Initially a power system model \citep{fodstad2022next}, it now includes other carriers such as heat, natural gas, and hydrogen \citep{backe2021heat,durakovic2024decarbonizing, egging2021freedom}.

The starting point of this study is the model presented in \citet{durakovic2024decarbonizing}. Several attributes of this model make it suitable to answer the research questions, as seen below. This model is suitable for answering the research questions due to several attributes. Its linear nature allows for high spatial resolution, representing 31 European countries while remaining computationally manageable. Additionally, EMPIRE is a multi-horizon stochastic program that accounts for hourly operational uncertainty within a long-term planning problem without excessive complexity. This enhances the study's robustness and reliability, especially for green hydrogen, which depends on weather-dependent \gls{vres} production.

The extended version of EMPIRE will be used on two case studies designed to answer the following research questions: 
\begin{enumerate}
    \item What are the upper bound additional cost estimates and renewable energy investments required to ensure that all new electrolysis plants produce green hydrogen in the European energy system from 2024 to 2048?
    \item How does each requirement for green hydrogen impact investment decisions and costs, being:
    \begin{enumerate}
        \item[(A)] additionality
        \item[(S)] spatial correlation
        \item[(T)] temporal correlation
        \item[(90)] renewable grid exemption
    \end{enumerate}
\end{enumerate}

The structure of the paper is as follows: \autoref{ch:lit} reviews related research to identify gaps and contributions. \autoref{ch:method} details the extended model structure and mathematical formulation. \autoref{ch:data} describes the case studies and data for model input parameters. Finally, \autoref{ch:results} presents and discusses the results, with conclusions in \autoref{ch:conclusion}.

\section{Literature review}\label{ch:lit}
The literature review first details the \gls{eu}'s definition of green hydrogen. It then discusses the extent to which this definition has been incorporated into previous energy system modeling studies, identifying existing research shortcomings.

\subsection{Green hydrogen definition}
\label{sec:green_def}

Developing the current definition of green hydrogen has been a prolonged process. A milestone was achieved with the Renewable Energy Directive (2009/28/EC) \citep{RED2009}, aiming to shift the \gls{eu} towards a sustainable, renewable-based energy system with binding \gls{vres} targets for 2020. The Revised Renewable Energy Directive (Directive (EU) 2018/2001) \citep{REDII2018}, or RED II, adopted in 2018, aligns with the Paris Agreement \citep{UNFCCC2015} and sets more ambitious targets for 2030 and 2050. RED II includes sector-specific regulations and introduces renewable fuels of non-biological origin (\gls{rfnbo}), under which green hydrogen is classified.

To extend RED II, the European Commission published two delegated acts supporting the \gls{eu} definition of green hydrogen \citep{euRenewableHydrogen2023}. Enacted into \gls{eu} law on June 20, 2023, these acts are crucial for the anticipated uptake of European hydrogen production. Clear definitions provide legal certainty for producers and consumers, mitigating risk and ensuring applicability for future subsidy schemes.

An increase in green hydrogen production will raise Europe's total power demand, necessitating truly renewable power sources. The Delegated Acts on Renewable Hydrogen complement the definition of green hydrogen by specifying additional power source requirements \citep{euRegulation2023R1184}.

Both on-grid and off-grid installations for electrolysis are considered for future investments, but this study focuses on grid-connected electrolysis. The Delegated Acts on Renewable Hydrogen specify criteria for ensuring green hydrogen production from grid-connected electrolysis, including:

\begin{enumerate}
    \item \textbf{Additionality}: Renewable power must come from installations started within 36 months before the hydrogen plant's launch and not part of pre-existing plans.
    \item \textbf{Spatial correlation}: The renewable energy source and hydrogen plant must be in the same power market bidding zone or a connected offshore zone.
    \item \textbf{Temporal correlation}: Hydrogen must be produced within the same hour as the renewable power it consumes.
\end{enumerate}

The requirements also include an exemption for bidding zones where the grid averages over 90\% renewable power, deeming all grid power as renewable for green hydrogen production. However, electrolysis power use is limited to a pre-decided proportion of total grid power. For additionality, countries with grid carbon intensity below $64.8$ gCO2/kWh are exempt from hydrogen-dedicated renewable capacity investments but must still ensure spatial and temporal correlation \citep{zeyen2024temporal}.

The new requirements have received mixed reactions; while they provide clarity, some stakeholders argue they are overly stringent. \citet{brandt2024cost} concluded that these regulations increase production costs, limiting green hydrogen expansion and delaying economies of scale \citep{HydrogenEuropeAdditionality2023}. Nonetheless, the regulations aim to ensure a renewable hydrogen sector and create a sustainable energy system. A clear definition of green hydrogen is essential to incentivize investments through future support mechanisms, including the European Hydrogen Bank mentioned in \autoref{ch:intro}.

\subsection{Green hydrogen definition in European energy system modeling}

\begin{table*}[H]
\caption{Model comparison for related literature.}\label{tab: model_comparison}
\centering
\small
\begin{tabular}{llllll}
\hline 
\textbf{Article} & \textbf{Year} & \textbf{Model} & \textbf{Temporal} & \textbf{Spatial} & \textbf{Electrolyser power}\\
&  && \textbf{resolution}& \textbf{resolution}& \textbf{source requirements}\\
\hline
\citet{seck2021hydrogen4eu} & 2021 & MIRET-EU & 12 hr & 27 nodes & No \\
\citet{arduin2022energy} & 2022 & Artelys Crystal Super Grid & 1 hr & 35 nodes & No \\
\citet{kountouris2023unified} & 2023 & Balmorel & 1 hr & 41 nodes & No \\
\citet{neumann2023potential} & 2023 & PyPSA-Eur-Sec & 3 hr & 181 nodes & No \\
\citet{durakovic2024decarbonizing} & 2024 & EMPIRE & 1 hr & 51 nodes & No \\
\citet{zeyen2024temporal} & 2024 & PyPSA-Eur & 1 hr & - & Green hydrogen compliant \\
\hline
\end{tabular}
\end{table*}

Multiple studies have used multi-carrier energy system models to explore hydrogen's role in Europe's green transition. Recent research distinguishes between hydrogen production methods, aiming to find hydrogen's role in a net-zero European energy system by 2050, as shown in \autoref{tab: model_comparison}. Despite the lack of explicitly considering policies, some models align with the \gls{eu} definition of green hydrogen. Using the Model of International Renewable Energy Transmission - European Union (MIRET-EU), \citet{seck2021hydrogen4eu} define green hydrogen as electrolysis-based without specific power source requirements. Results indicate most hydrogen production will be from off-grid electrolyzers with renewable sources. The Baltic Model of Regional Electricity Liberalization (balmorel) by \citet{kountouris2023unified} shows hourly hydrogen production follows renewable energy availability, but lacks geographical correlation. \citet{neumann2023potential} use Python for Power System Analysis - Europe - Sector Coupling (PyPSA-Eur-Sec) to show that enforcing the new European definition of green hydrogen is feasible with a manageable cost premium. \citet{durakovic2024decarbonizing} use the energy system optimization model EMPIRE to study hydrogen uptake in the absence of Russian gas, labeling all electrolysis as green without power source requirements. Despite the lack of formal definitions, \citet{durakovic2024decarbonizing} contributes to the green hydrogen pathway by accounting for uncertainty in power demand and \gls{vres} production profiles.

In contrast, \citet{zeyen2024temporal} aligns with the \gls{eu} green definition, using PyPSA-Eur to investigate the effects of enforcing different green hydrogen production standards. The study compares the new European definition to more flexible requirements regarding the temporal and spatial correlation of \gls{vres} and electrolysis. They conclude that flexible electrolysis operation should be supported to ensure low emissions and competitive production costs. However, the study only considers two individual years (2025 and 2030) and focuses on one country at a time (Germany or the Netherlands).

Current literature on Europe's future energy system has shortcomings in assessing the impact of green hydrogen requirements. Models implementing the European definition of green hydrogen often simplify complexity by limiting spatial resolution or energy carriers and sector coupling. There is a need to investigate the effects of enforcing the European definition in a complex energy system model with a European scope.

This study models a very restrictive scenario, where all regulatory criteria are met immediately, without phase-in periods. This approach examines the upper-bound impacts on system costs and assesses how the hydrogen production system will respond to stringent regulations. Additionally, investigating these regulations under short-term uncertainty regarding weather conditions and electricity demand is beneficial, as it captures the challenges in green hydrogen production powered by \gls{vres}.

\citet{zeyen2024temporal} conclude that stricter short-term rules, such as hourly matching until renewable targets are met and coal is phased out, will limit emissions. Incorporating the 90\% renewable grid exemption in the analysis reflects easing the hourly matching requirement during the later stages of the period examined in this article.

\section{Methodology} \label{ch:method}
This section outlines the mathematical model used to address the research questions from \autoref{ch:intro}, beginning with a qualitative description of EMPIRE, followed by an explanation of the new model implementations.

\subsection{Model structure: EMPIRE’s multi-horizon stochastic optimization for energy markets}

This paper uses the multi-horizon, stochastic optimization model EMPIRE \citep{backe2022empire} to study the effects of aligning hydrogen production with the \gls{eu} definition of green hydrogen. EMPIRE is a multi-carrier energy system model for operations and investment decisions in the European energy market. The system is modeled as a connected graph of nodes and arcs, where nodes represent markets with exogenous and endogenous demand for multiple energy carriers and industries, and arcs represent the exchange of commodities between nodes. EMPIRE includes two temporal scales: long-term strategic 3-year periods and short-term operational hours, forming a linear multi-horizon stochastic program. First-stage decisions involve investments in each long-term period, while second-stage decisions cover operations in several stochastic scenarios within each investment period.

The distinction between first-stage and second-stage decisions is crucial for understanding how uncertainty impacts the stochastic model results. There is only one consistent first-stage investment decision per technology, node, and investment period. In contrast, there are several second-stage decisions per technology and node, even for the same hour, season, and investment period. This means investments are made with the understanding that hourly realizations of uncertain parameters are not known with certainty and could vary across different stochastic scenarios.

Stochastic scenarios are generated through random sampling of representative operational time windows, as described in \citet{backe2021stable}, to represent short-term uncertainty. The stochastic parameters include: exogenous power demand profiles in each node at each operational hour, \gls{vres} production availability at each operational hour, and reservoir hydropower availability on a seasonal basis.

The model does not consider every hour of the year due to computational complexity, but hourly resolution is crucial for simulating operational dynamics. To balance this, representative time periods with hourly resolution are used. Each period lasts one week to capture daily variations while maintaining tractable complexity. Multiple realizations of these weeks reflect uncertainty. This approach's value is further discussed in \citet{backe2021stable}.

\begin{figure*}[t]
    \centering
    \includegraphics[width=1\linewidth]{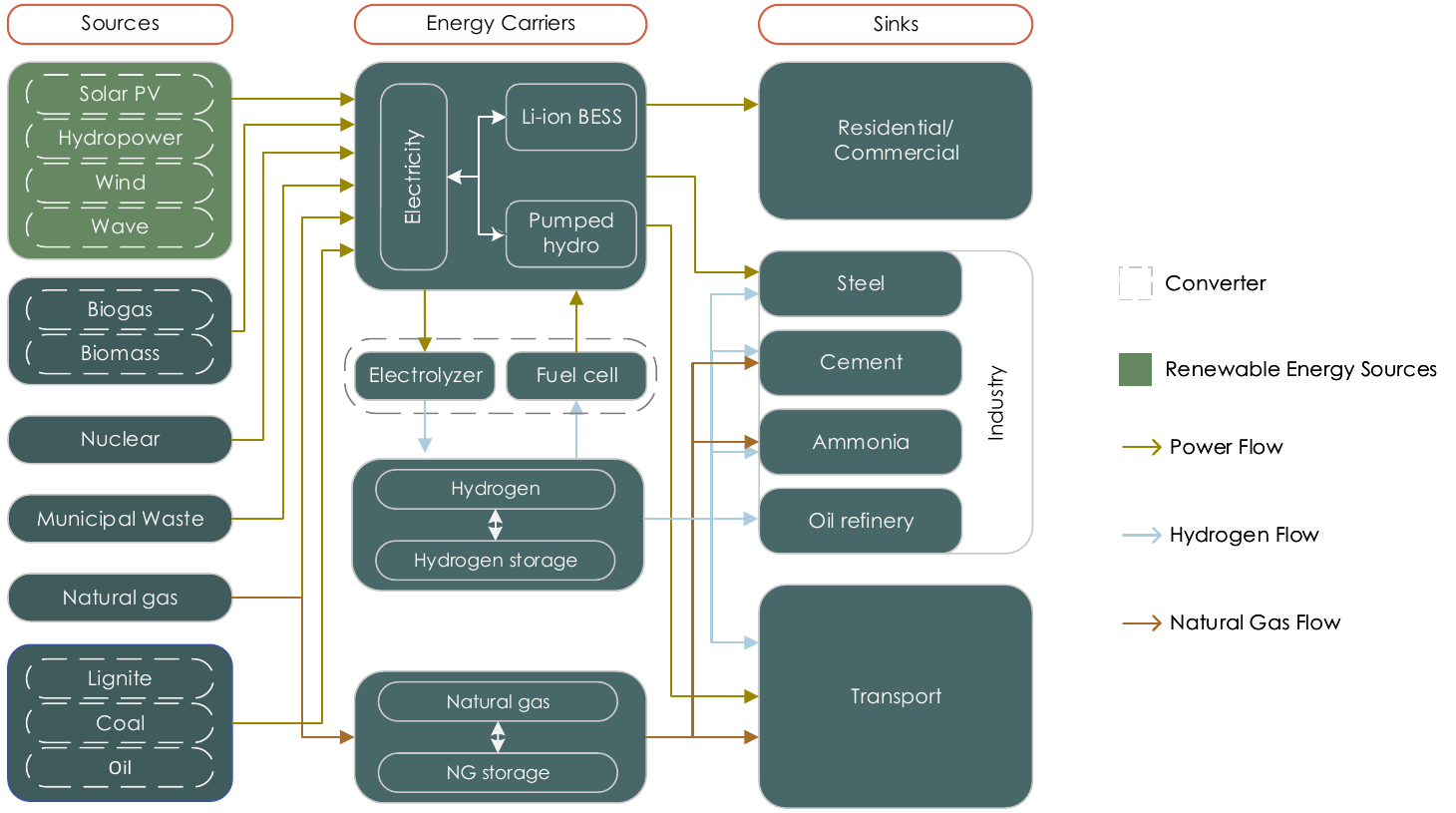}
    \caption{Graphical representation of energy flows in EMPIRE between sources and sinks.}
    \label{fig:powersys}
\end{figure*}

EMPIRE finds the cost-optimal solution for long-term energy system planning by minimizing total system costs. Production and storage capacities are invested for each node, constrained by maximum capacity expansion per period and total capacity for the model horizon. Transmission between nodes is considered only for energy carriers and \gls{co2}, also constrained by maximum expansion and total capacity. Second-stage decisions are limited by investment decisions, up-ramping limitations, and asset availability to meet hourly demands.

\autoref{fig:powersys} displays the energy flow in the system with the following definitions:
\begin{itemize}
    \item Sources: Production assets of energy carriers, which can be stored, converted, or used to meet demand.
    \item Energy carriers: Intermediates between energy sources and consumption.
    \item Sinks: Market sectors demanding energy from different carriers.
\end{itemize}

\subsection{Mathematical formulation: enhancing EMPIRE for green hydrogen and pipeline repurposing}
This section presents the sets, parameters, variables, and constraints of the extensions made to EMPIRE in this work. The contribution to the model is divided into two parts: enforcing electrolysis to comply with the European green hydrogen definition and allowing natural gas pipelines to be repurposed for hydrogen. For a complete mathematical formulation of the model, see \autoref{appendix_a}.

\subsubsection{Green hydrogen production}
New sets, variables, and parameters have been added to require that electrolysis qualify as green. $\mathcal{G}^{VRES}$ represents all generators categorized as \gls{vres} and is a subset of all power generators. This set of generators is approved to power electrolysis for green hydrogen production.  $\mathcal{N}^{90}_i$ represents all nodes that qualify for the 90\% renewable grid exemption in period $i$.

\textbf{Additionality: }
The additionality constraint, shown in \autoref{eq:additionality_investment}, limits electrolyzer investments $x^e_{n,i}$ within the set $\mathcal{E}$ to the generator capacity built $x^{g}_{n,i}$ within the set of \gls{vres} generators $\mathcal{G}^{VRES}$ for each period $i$ and node $n$. The exogenous parameter $\eta^{PW}_e$ represents the constant power (PW) consumption for producing one ton of hydrogen (\gls{h2}) at electrolyzer $e$, ensuring that electrolysis power demand does not exceed the capacity of newly built \gls{vres} generators.

\autoref{eq:additionality_investment} guarantees a minimum of new \gls{vres} investments per electrolyzer, acknowledging that \gls{vres} electricity is not available 100\% of the time. This simplification avoids excessive \gls{vres} investments, considering flexible electrolyzer operation. However, \autoref{eq:additionality_investment} alone does not ensure additionality as defined by \citet{euRenewableHydrogen2023} unless electrolyzers perfectly adapt to \gls{vres} output. It constrains investment timing, not production volumes.

No exemptions from \autoref{eq:additionality_investment} are considered, even though countries with grid carbon intensity below $64.8$ gCO2/kWh are exempt \citep{zeyen2024temporal}. This stricter implementation triggers more renewable investments than in reality.

\small
\begin{align}\label{eq:additionality_investment}
    \begin{split}
       \sum_{e \in \mathcal{E}}(x^e_{n,i} \times \eta^{PW}_e) \leq \sum_{g \in \mathcal{G}^{VRES}} x^{g}_{n,i}  \quad \\ \forall ~n \in \mathcal{N} \setminus \mathcal{N}^{90}_i, ~i \in \mathcal{I}. 
    \end{split}
\end{align}
\normalsize

\textbf{Spatial and temporal correlation: }\autoref{eq:green_power_cap} and \autoref{eq:power_to_H2} define the spatial and temporal correlation between renewable power generation and electrolysis. \autoref{eq:green_power_cap} limits the hourly power for electrolysis $y^{PW4H_2}_{n,h,i,\omega}$ to the total available power from additional \gls{vres}. Parameter $\alpha_{g,n,h,i,\omega}$ indicates the stochastic availability of a renewable source, based on historical data, while $i^{life}_{g}$ accounts for generator depreciation.

Unlike the \gls{eu}'s monthly matching rule until 2030 \citep{euRenewableHydrogen2023}, \autoref{eq:green_power_cap} requires stricter hourly matching, necessitating more flexibility from storage and electrolyzer operations before 2030.

\small
\begin{align}\label{eq:green_power_cap}
    \begin{split}
    y^{PW4H_2}_{n,h,i,\omega}\leq \sum_{g \in \mathcal{G}^{VRES}}(\alpha_{g,n,h,i,\omega} \times \sum_{j=i'}^{i}x^g_{n,j})
    \quad  \\
    \forall ~n \in \mathcal{N} \setminus \mathcal{N}^{90}_i, ~i \in \mathcal{I},~h \in \mathcal{H},\\
    ~\omega \in \Omega,~i'=\max\{1,i-i^{life}_{g}\}.
    \end{split}
\end{align}
\normalsize

\autoref{eq:power_to_H2} ensures that the total power consumed by electrolysis does not exceed the available green power in each node for each operational hour in each scenario.

\small
\begin{align}\label{eq:power_to_H2}
    \begin{split}
    \sum_{e \in \mathcal{E}} (y^{H_2,source}_{e,n,h,i,\omega} \times \eta^{PW}_e) \leq y^{PW4H_2}_{n,h,i,\omega} \\ 
    \forall ~e \in \mathcal{E},~n \in \mathcal{N} \setminus \mathcal{N}^{90}_i,
    ~i \in \mathcal{I},~h \in \mathcal{H},~\omega \in \Omega. 
    \end{split}
\end{align}
\normalsize

\textbf{90\% renewable grid exemption: }
\autoref{eq:renewable_grid_exemption} ensures that nodes exempt from green hydrogen production requirements maintain over 90\% renewable power production in the respective periods. This is achieved by requiring that, for all seasons $\mathcal{S}$ and hours $\mathcal{H}^{s}$, representing annual power production, more than 90\% of total power production in that node and period comes from \gls{vres} generators $\mathcal{G}^{VRES}$.

\small
\begin{align}\label{eq:renewable_grid_exemption}
    \begin{split}
        0.9 \times \sum_{s \in \mathcal{S}} \sum_{h \in \mathcal{H}^s} \sum_{g \in \mathcal{G}} y^{PW,source}_{g,n,h,i,\omega} \leq
        \sum_{s \in \mathcal{S}} \sum_{h \in \mathcal{H}^s} \sum_{g \in \mathcal{G}^{VRES}} y^{PW,source}_{g,n,h,i,\omega}\\
        \forall ~n \in \mathcal{N}^{90}_i, ~i \in \mathcal{I},~\omega \in \Omega. 
    \end{split}
\end{align}
\normalsize

\subsubsection{Repurposing natural gas pipelines}
The network of transmitting commodities is modeled as bidirectional arcs between two nodes. $\mathcal{L}^{c}_n$ is the set of all possible bidirectional arcs to node $n$ for commodity $c$. The model does not allow investments in natural gas pipelines and assumes no depreciation, so the initial capacity $\bar{x}^{NG Pipeline}_{n,m}$ remains constant across all investment periods.

\autoref{repurpose_cost} extends the objective function to include the costs of repurposing \gls{ng} pipelines. The repurposing cost is a factor $\kappa^{repurpose}$ of the cost of building new hydrogen pipelines with equal capacity. Pipeline capacity is represented by the reduction in natural gas capacity $x^{{repurpose}}_{n,m,i}$ times the flow factor $\eta^{repurpose}$ for converting natural gas to hydrogen. The cost of new hydrogen pipelines is $c^{H_2Pipeline}_{i}$, discounted by $(1+r)^{L^{period}(i-1)}$ with a discount rate $r$.

\small
\begin{align}\label{repurpose_cost}
    \begin{split}
         \kappa^{repurpose} \times \sum_{i \in \mathcal{I}} \sum_{n \in \mathcal{N}} \sum_{m \in \mathcal{L}^{NG}_n} [(1+r)^{L^{period}(i-1)} \times\\ c^{H_2Pipeline}_{i} \times \eta^{repurpose} \times x^{{repurpose}}_{n,m,i}] 
    \end{split}
\end{align}
\normalsize

\autoref{eq:trans_NG_cap} ensures that the capacity of natural gas pipelines $v^{NG Pipeline}_{n,m,i}$ between two nodes is equal to the initial capacity $\bar{x}^{NG Pipeline}_{n,m}$ reduced by the capacity that has been repurposed $x^{{repurpose}}_{n,m,i}$ up to and including the current investment period.

\small
\begin{align}\label{eq:trans_NG_cap}
    \begin{split}
       v^{NG Pipeline}_{n,m,i} = \bar{x}^{NG Pipeline}_{n,m} - \sum_{j=1}^i x^{{repurpose}}_{n,m,j}\\ \forall ~n \in \mathcal{N},~m \in \mathcal{L}^{NG}_n,~i \in \mathcal{I}.
    \end{split}
\end{align}
\normalsize

\autoref{eq:max_repurpose} limits the total capacity repurposed for all investment periods to the initial capacity of the natural gas pipelines.

\small
\begin{align}\label{eq:max_repurpose}
    \begin{split}
       \sum_{i \in \mathcal{I}} x^{{repurpose}}_{n,m,i} \leq\bar{x}^{NG Pipeline}_{n,m} \\
       \forall ~n \in \mathcal{N},~m \in \mathcal{L}^{NG}_n.
    \end{split}
\end{align}
\normalsize

Finally, \autoref{eq:trans_h2_cap} adds repurposed pipelines to hydrogen pipeline capacities. The total hydrogen pipeline capacity $v^{H_2Pipeline}_{n,m,i}$ includes new hydrogen pipelines $x^{H_2Pipeline}_{n,m,i}$, repurposed natural gas pipelines $x^{repurpose}_{n,m,i}$, and initial capacity $\bar{x}^{H_2Pipeline}_{n,m,i}$. Initial capacity for each period is a percentage of the initial capacity in the first investment period, accounting for depreciation. Parameter $\eta^{repurpose}$ is the flow factor for repurposing from natural gas to hydrogen, with $x^{repurpose}_{n,m,i}$ representing lost natural gas capacity. Parameter $i^{life}_{H_2Pipeline}$ represents asset lifetime and accounts for hydrogen pipeline depreciation.

\small
\begin{align}\label{eq:trans_h2_cap}
    \begin{split}
        v^{H_2Pipeline}_{n,m,i} = \bar x^{H_2Pipeline}_{n,m,i} + \\
        \sum_{j=i'}^i (x^{H_2Pipeline}_{n,m,j} + \eta^{repurpose} \times x^{repurpose}_{n,m,j}) \\
        \forall ~n \in \mathcal{N},~m \in \mathcal{L}^{NG}_n,~i \in \mathcal{I}, \\
        ~i'=\max\{1,i-i^{life}_{H_2Pipeline}\}.
    \end{split}
\end{align}
\normalsize

\section{Case studies and data} \label{ch:data}
This section describes the case study structure and presents the assumptions and sources for the input to the model. More detailed descriptions of the latter can be found in \citet{durakovic2024decarbonizing}.

\begin{table*}
\caption{Fixed annual hydrogen production quantities per period.}\label{tab: minimum_hydrogen}
\centering
\small
\begin{tabular}{lllllllll}
\hline
& \textbf{2024-} & \textbf{2027-} & \textbf{2030-} & \textbf{2033-} & \textbf{2036-} & \textbf{2039-} & \textbf{2042-} & \textbf{2045-} \\
& \textbf{2027} & \textbf{2030} & \textbf{2033} & \textbf{2036} & \textbf{2039} & \textbf{2042} & \textbf{2045} & \textbf{2048} \\
\hline
Fixed hydrogen & & & & & & & & \\
production [Mt/yr] & 10 & 10 & 10 & 11 & 11 & 12 & 15 & 15 \\
\hline
\end{tabular}
\end{table*}

\subsection{Evaluating green hydrogen production requirements}
\label{subsec: case_studies}

This paper runs six cases to answer the research questions in \autoref{ch:intro}. The cases differ in power supply requirements for electrolysis, which is the only form of hydrogen production considered, excluding hydrogen production from natural gas reformers.

To see the isolated impact of the different requirements for green hydrogen, the same hydrogen production is required over all six cases using \autoref{eq:prod_LB}:

\small
\begin{align}\label{eq:prod_LB}
    \begin{split}
        \sum_{e \in \mathcal{E}} \sum_{n \in \mathcal{N}} \sum_{h \in \mathcal{H}}  y^{H_2,source}_{e,n,h,i,\omega} = H_i \quad \\
        \forall ~i \in \mathcal{I},~\omega \in \Omega.
    \end{split}
\end{align}
\normalsize

\autoref{eq:prod_LB} requires each case to produce the hydrogen quantity $H_i$ in every stochastic scenario, uniquely defined for each period. The hydrogen flow variable $y^{H_2,source}_{e,n,h,i,\omega}$ is given for electrolyzer type $e \in \mathcal{E}$. The hydrogen production quantities $H_i$ match the levels chosen by EMPIRE in \citet{durakovic2024decarbonizing} and shown in \autoref{tab: minimum_hydrogen}. These numbers align with future hydrogen demand projections for Europe by \citet{EuropeanHydrogenObservatory2024}, though short-term demand is high and long-term demand is low.

To answer the first research question, two cases are run to quantify the cost of green hydrogen. The first case, Base, envisions a Europe without requirements for the source of power supplied to electrolysis. The second case, AST90, requires all electrolysis to adhere to the \gls{eu}'s criteria for green hydrogen. The name AST90 reflects the requirements for green hydrogen: additionality (A), spatial correlation (S), and temporal correlation (T), including the exemption for nodes where 90\% of the grid is renewable (90).

To answer the second research question, the AST90 case is compared to four cases named ST90, AT90, AS90, and AST. Definitions are provided below. In all new cases, one of the requirements, or the renewable grid exemption, is removed as a condition to qualify for green hydrogen. What rules apply in the different cases are displayed in \autoref{tab: case_rules}. The model changes for each case as follows:

\begin{table*}[ht]
\caption{Green hydrogen criteria for different cases.}
\label{tab: case_rules}
\centering
\small
\begin{tabular}{lllllll}
\hline
\textbf{Item} & \textbf{Base} & \textbf{AST90} & \textbf{ST90} & \textbf{AT90} & \textbf{AS90} & \textbf{AST} \\
\hline
Additionality & & X &  & X & X & X \\
Spatial correlation & & X & X &  & X & X \\
Temporal correlation & & X & X & X &  & X \\
90\% exemption & & X & X & X & X &  \\
\hline
\end{tabular}
\end{table*}

\textbf{ST90: } 
\autoref{eq:additionality_investment}, which limits the investments in electrolyzer capacity to the \gls{vres} capacity built in the same investment period, is removed from the model. \autoref{eq:green_power_cap} is replaced by \autoref{eq:green_power_cap_no_additivity} so that power for electrolysis is limited by all \gls{vres} generator capacities $v^g_{n,i}$, in contrast to just newly built capacities $x^g_{n,i}$.

\small
\begin{align}\label{eq:green_power_cap_no_additivity}
    \begin{split}
        y^{PW4H_2}_{n,h,i,\omega}\leq \sum_{g \in \mathcal{G}^{VRES}}(\alpha_{g,n,h,i,\omega} \times v^g_{n,i})\\
         \forall ~n \in \mathcal{N} \setminus \mathcal{N}^{90}_i, ~i \in \mathcal{I},~h \in \mathcal{H},
        ~\omega \in \Omega.
    \end{split}
\end{align}
\normalsize

\textbf{AT90: }
\autoref{eq:green_power_cap} is replaced by \autoref{eq:green_power_cap_no_spatial} so that the sum of power to electrolysis $y^{PW4H_2}_{n,h,i,\omega}$ is limited by the sum of available power production from additional \gls{vres} $x^g_{n,i}$, over all nodes that do not qualify for the renewable grid exemption in the respective period.

\small
\begin{align}\label{eq:green_power_cap_no_spatial}
    \begin{split}
    \sum_{i \in \mathcal{I}}\sum_{n \in \mathcal{N} \setminus \mathcal{N}^{90}_i} y^{PW4H_2}_{n,h,i,\omega}\leq \\
    \sum_{g \in \mathcal{G}^{VRES}}\sum_{i \in \mathcal{I}}\sum_{n \in \mathcal{N} \setminus \mathcal{N}^{90}_i}(\alpha_{g,n,h,i,\omega} \times \sum_{j=i'}^{i}x^g_{n,i})
    \quad \\
    \forall ~h \in \mathcal{H},
    ~\omega \in \Omega, 
    ~i'=\max\{1,i-i^{life}_{g}\}.
    \end{split}
\end{align}
\normalsize

\textbf{AS90: }
\autoref{eq:green_power_cap} is replaced by \autoref{eq:green_power_cap_no_temporal} so that the sum of power to electrolysis $y^{PW4H_2}_{n,h,i,\omega}$ is limited by the sum of available power production from additional \gls{vres} $x^g_{n,i}$, for all operational hours in an investment period. This represents a shift from hourly to annual matching in the temporal correlation.

\small
\begin{align}\label{eq:green_power_cap_no_temporal}
    \begin{split}
    \sum_{s \in \mathcal{S}} \sum_{h \in \mathcal{H}^s} y^{PW4H_2}_{n,h,i,\omega}\leq \sum_{s \in \mathcal{S}} \sum_{h \in \mathcal{H}^s} \sum_{g \in \mathcal{G}^{VRES}}(\alpha_{g,n,h,i,\omega} \times \sum_{j=i'}^{i}x^g_{n,i})
    \quad \\ \forall ~n \in \mathcal{N} \setminus \mathcal{N}^{90}_i, ~i \in \mathcal{I},~\omega \in \Omega,~i'=\max\{1,i-i^{life}_{g}\}.
    \end{split}
\end{align}
\normalsize

\textbf{AST: } The set of countries qualifying for the renewable grid exemption $\mathcal{N}^{90}_i$ is modified to be empty, meaning no countries are exempt from green hydrogen requirements. Therefore, \autoref{eq:additionality_investment}, \autoref{eq:green_power_cap}, and \autoref{eq:power_to_H2} apply to all nodes $n \in \mathcal{N}$ in all investment periods $i \in \mathcal{I}$, and \autoref{eq:renewable_grid_exemption} is removed.

\subsection{Data}\label{sec:data}

EMPIRE minimizes total system costs, categorized by investment and operational costs, all adjusted with a 5\% discount rate as per \citet{skar2014future, backe2022empire}, resulting in costs being considered in net present values for the year 2024. The cost of investing in increased capacities is aligned with \citet{seck2021hydrogen4eu}. 

The model includes 52 nodes across Europe: 30 for European countries, 5 for Norway's power price zones, 14 offshore wind farm nodes, and a single offshore energy hub node. It also features two nodes for the Sleipner and Draupner offshore platforms, part of the North Sea's natural gas network. Initial power generator and storage capacities are based on \citet{ENTSOE2022}. Hourly data for stochastic parameters (2015-2019) are from renewables.ninja \citep{pfenninger2016long,staffell2016using} and load and hydropower data from ENTSO-E Transparency Platform \citep{hirth2018entso}.

Investment periods are 3 years to accommodate the additionality requirement for green hydrogen, namely that investments in \gls{vres} and electrolyzers are made within 36 months. The total horizon spans eight investment periods from 2024 to 2048.

Renewable sources for green electrolysis is aligned with the \gls{eu} definition of \gls{rfnbo} \citep{EU2023RFNBO}, namely wave, geothermal, solar, hydropower, and wind.

To qualify for the renewable grid exemption, at least 90\% of annual power generation in a node must come from \gls{vres}. \autoref{tab:90percentshare_vres} in \autoref{appendix_b} provides an overview of countries qualifying for each investment period. This pre-defined set maintains a linear model structure, based on power production results from EMPIRE as presented in \citet{durakovic2024decarbonizing}.

The cost of repurposing natural gas pipelines for hydrogen is assumed to be 25\% of the cost of building new hydrogen pipelines with equal capacities \citep{DNV2023}. Repurposed pipelines are assumed to have 80\% of the energy flow capacity of natural gas pipelines \citep{ACER2021}. Investment costs for hydrogen pipelines, electrolyzers, and reformers, along with future cost trajectories, are sourced from \citet{seck2021hydrogen4eu}.

Natural gas production and transmission parameters follow \citet{durakovic2024decarbonizing}. Power, natural gas, and hydrogen within EMPIRE face exogenous and endogenous demands, including transport fuel use across nodes and periods. These demands align with \gls{eu} Reference Scenarios and node-specific data \citep{EUReferenceScenario2020, OdysseeMureEfficiency2023, EUReferenceScenario2016}. Europe's overall power demand is projected to increase by 18.7\% from 2024-2027 to 2045-2048. Unmet power demand incurs a penalty of €22,000 per MWh \citep{VoLL2013}.

Industry demands for steel, cement, ammonia, and oil refining align with projections from authoritative sources \citep{EPRS2021, egenhofer2014final, usgs2016, cement2018technology, iea2022world}. Future growth of steel demand is assumed to follow the trajectories of the International Energy Agency \citep{IEA2020IronSteelRoadmap}. 

Industrial processes in the model create endogenous demands for all energy commodities, especially hydrogen and natural gas. Endogenous power demand is driven by electrolysis for hydrogen production, initially consuming 57.5 MWh per ton of hydrogen, decreasing with technological advancements \citep{H2TPC2020}. Remaining natural gas and hydrogen demand fulfills power and industrial sector needs. The size of endogenous demand, particularly for power generation using natural gas and hydrogen, depends on generator efficiency levels, with data from \citet{ASSET2018}.

The model limits carbon dioxide (\gls{co2})  emissions for each investment period by imposing a cap on the overall emissions within the system across all sectors, adhering to the European Commission objectives \citep{EuropeanCommission2018} and the \gls{euets}. \gls{co2} emission intensities are derived from \citet{IPCC2006} for power generators and \citet{HYBRIT2020}, \citet{IEA2018Cement}, and \citet{Yara2024AmmoniaUreaCost} for the industrial sector. 

The model considers an explicit representation of a \gls{ccs} network featuring \gls{co2}  pipelines and storage sites as presented by \citet{durakovic2024decarbonizing}. This network captures \gls{co2} from power generators, reformers, selected industries, pipeline transport, and permanent storage in the North Sea. \gls{co2}  capture rates are sourced from \citet{IPCC2006}, \citet{seck2021hydrogen4eu}, \citet{IEAGHG2013Steel} and \citep{ETSAP2010Cement} for power generators, reformers, steel, and cement, respectively.

Note that all future data assumptions are subject to uncertainty, impacting results if projections are incorrect. Although EMPIRE is a stochastic model, assumptions about costs, technology characteristics, and annual demand are deterministic. All instances use the same future projections, and this paper focuses on differences between cases for the same future pathway. All input data assumptions are based on transparent sources, with full code and data available as open access on the public project Github page \citep{empire_green_hydrogen}.

\section{Results and discussion} \label{ch:results}

This section presents the results for the two research questions separately, illustrating cost and investment differences for the cases described. The only differences between the cases are the changes related to green hydrogen policies; all other data assumptions, including stochastic scenarios, remain the same.

\subsection{The cost of green hydrogen}
\begin{figure}[ht]
    \centering
    \includegraphics[width=7cm,height=10cm]{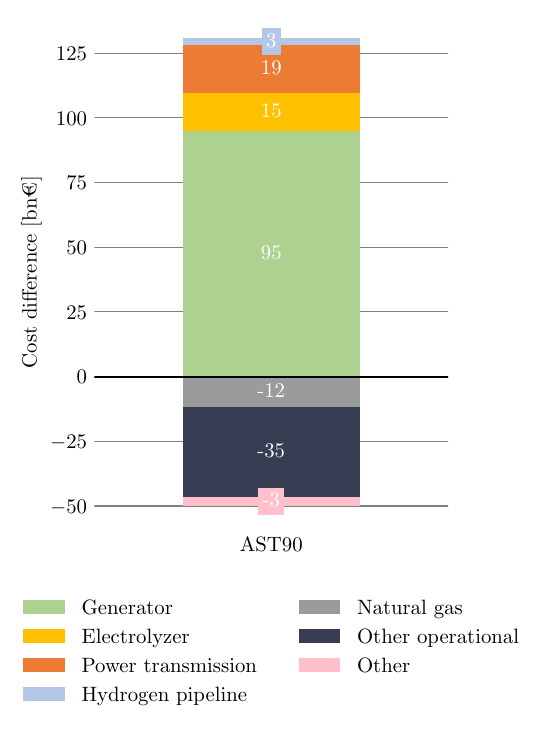}
    \caption{The expected cost difference of ensuring all future hydrogen from electrolyzers is green hydrogen by cost category. AST90 cost difference from Base.}
    \label{fig: OBJ_part1}
\end{figure}

An overall cost breakdown in \autoref{fig: OBJ_part1} illustrates the difference between Base and AST90. The total cost difference shows an increase of €82 billion from Base to AST90, representing a 3.7\% rise in overall system costs. The cost increase is most notable in the first investment periods (2024-2030) due to the renewable investments required for green hydrogen uptake.

\begin{figure*}[ht]
    \begin{multicols}{2}
        \begin{subfigure}[b]{0.45\textwidth}
            \includegraphics[width=\textwidth]{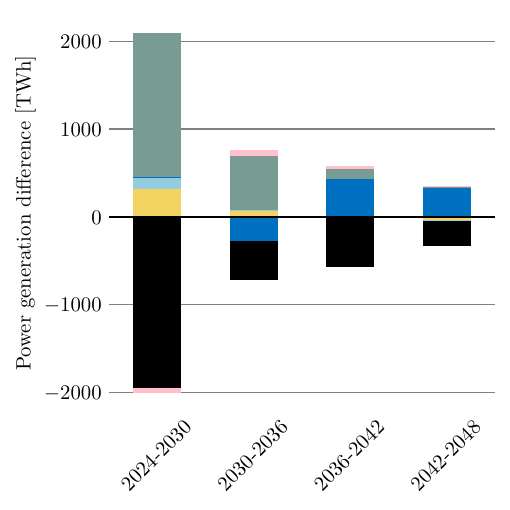}
            \caption{Total expected accumulated power generation difference for 6-year periods}
            \label{fig: gen_diff_part1}
        \end{subfigure}

        \begin{subfigure}[b]{0.45\textwidth}
            \includegraphics[width=\textwidth]{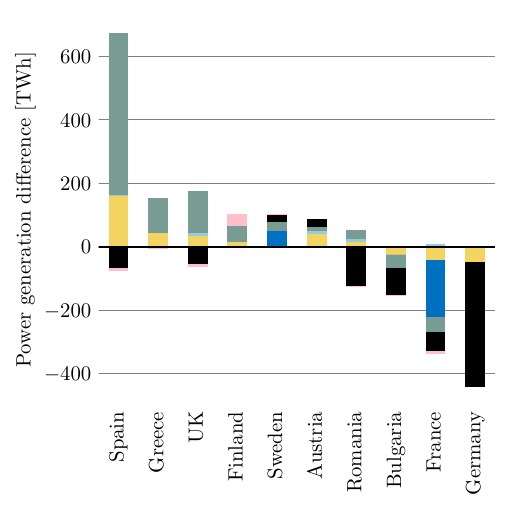}
            \caption{Geographical differences of total expected accumulated power generation between 2024 and 2048 for selected countries}
            \label{fig: gen_diff_spat_part1}
        \end{subfigure}        
    \end{multicols}
    \begin{subfigure}[b]{\textwidth}
        \includegraphics[width=\textwidth]{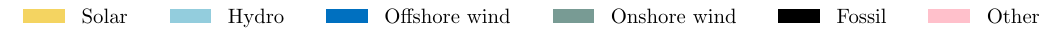}
    \end{subfigure}    
    \caption{Expected power generation difference from Base to AST90 across all stochastic scenarios.}
\end{figure*}

The primary cost variations stem from changes in capital expenditure for power generation, with increased investments in renewable generators, decreased natural gas usage, and reduced operational costs. These changes are largely due to reduced fossil fuel consumption in AST90 compared to the base.

The largest net cost increase is in power generation capital expenditure, amounting to €48 billion (2.1\% of total system costs). This increase is the difference between the €95 billion rise in power generator capital expenditure and the €47 billion reduction in expected operational expenditure. Given that power input costs are the largest component of the levelized cost of hydrogen \citep{IRENA2020}, this sector sees the highest cost increase under green hydrogen production.

The second largest cost component is capital expenditure on electrolyzers. These require more flexible operation and higher capital investments to achieve the same production levels as constant output, as they cannot run on fossil power and depend on variable power sources.

Increased hydrogen pipeline investments address the spatial and temporal inflexibility caused by these requirements. Higher power transmission costs complement increased \gls{vres} investments, as higher potential power peaks in one country necessitate the capacity to transmit this surplus to neighboring countries.

The following sections will analyze cost differences for expected power production, electrolyzers, and hydrogen and power transmission.

\subsubsection{Changes in expected power generation}
\autoref{fig: gen_diff_part1} shows the differences in expected power generation for AST90 and Base. The greatest differences occur in the first periods, increasing total expected production from \gls{vres} as fossil-based expected production is reduced. The findings suggest that committing to green hydrogen production from electrolyzers will boost renewable energy investments across Europe in the coming years. Fossil-based power production is expected to be 6\% lower between 2024 and 2027, reducing operational expenditure in AST90 compared to Base. Expected output from onshore wind, solar, and hydropower increase in the first investment periods, especially onshore wind. Differences in expected \gls{vres} output diminish in later periods (2036-2048), while expected offshore wind output increases in AST90 compared to Base.

\autoref{fig: gen_diff_spat_part1} presents the geographical differences in expected power generation from Base to AST90. Countries like Spain, Greece, and the United Kingdom see increased expected renewable power production, while centrally located countries like France and Germany see fewer investments in fossil and renewables, favoring supply from other countries via the upgraded transmission grid.

\subsubsection{Changes in electrolyzers}
\begin{figure*}[ht]
     \centering
     \begin{multicols}{2}
        \begin{subfigure}[b]{0.3\textwidth}
            \includegraphics[width=\textwidth]{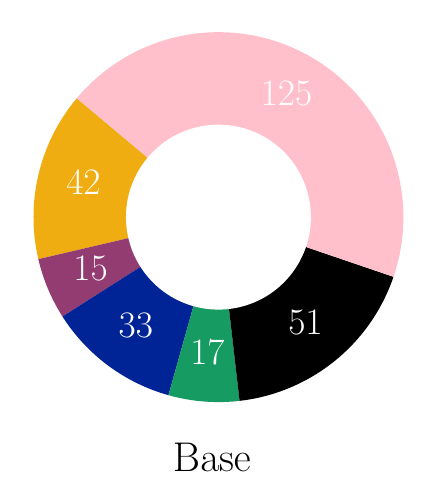}
        \end{subfigure}
        
        \begin{subfigure}[b]{0.3\textwidth}
            \includegraphics[width=\textwidth]{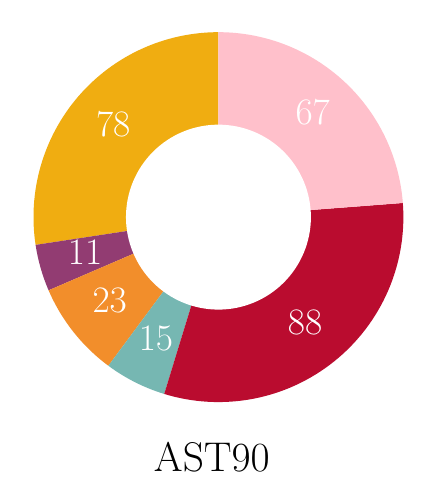}
        \end{subfigure}
    \end{multicols}
    \begin{subfigure}[b]{0.6\textwidth}
        \includegraphics[width=\textwidth]
        {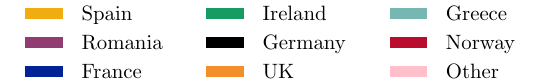}
    \end{subfigure}
    \caption{Total expected regional hydrogen production differences in Mton from 2024-2048 across all stochastic scenarios.}
    \label{fig: elyzer_pie}
\end{figure*}

As seen in \autoref{fig: OBJ_part1}, electrolyzer costs increase by €15 billion. This trend mirrors \gls{vres} investments, with higher electrolyzer capacity in AST90 compared to Base in the initial periods, followed by a decreasing difference in later years. Expected electrolyzer capacity factors are 90\% for Base and 77\% for AST90 in the first period, converging to 48\% in the final period for both cases. The future decrease in expected capacity factor is due to increased power production variability and reduced projected electrolyzer capital expenditure from economies of scale.

The increase in capital expenditure on electrolyzers is also due to the spatial inflexibility of the green hydrogen definition. Because of the spatial correlation requirement, a country's electrolyzers cannot import power from neighbors if there is a \gls{vres} production deficit in that country (see \autoref{sec:green_def}).

\autoref{fig: elyzer_pie} shows where hydrogen is expected to be produced in Base and AST90. In AST90, electrolyzer capacity is concentrated in fewer countries with competitive advantages in \gls{vres} production, such as Spain and Greece, due to their climate. Access to offshore wind is also crucial, with significant hydrogen production in Norway and the United Kingdom. Norway's exemption from green hydrogen production rules, due to over 90\% renewable power generation, makes it an attractive location for green hydrogen production.

\subsubsection{Hydrogen and power transmission upgrades}

\begin{figure}
    \centering
    \begin{subfigure}[b]{0.4\textwidth}
        \centering  
        \includegraphics[width=\textwidth]{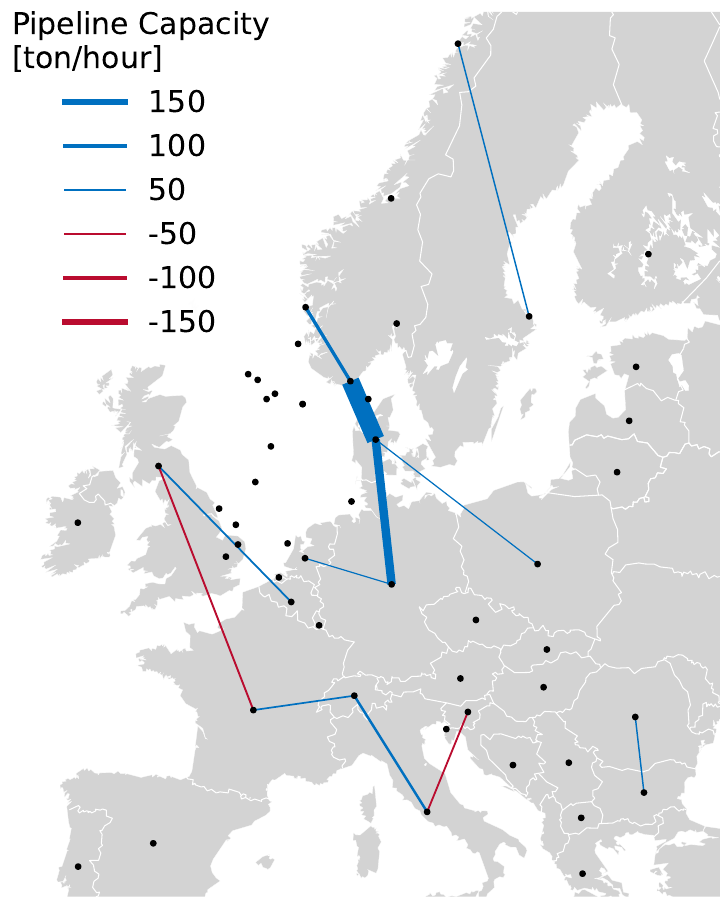}
        \caption{Hydrogen pipeline capacity difference more than 25 ton/hour.}
        \label{fig: pipeline}
    \end{subfigure}
    \hspace{0.05\textwidth}  
    \begin{subfigure}[b]{0.4\textwidth}
        \centering  
        \includegraphics[width=\textwidth]{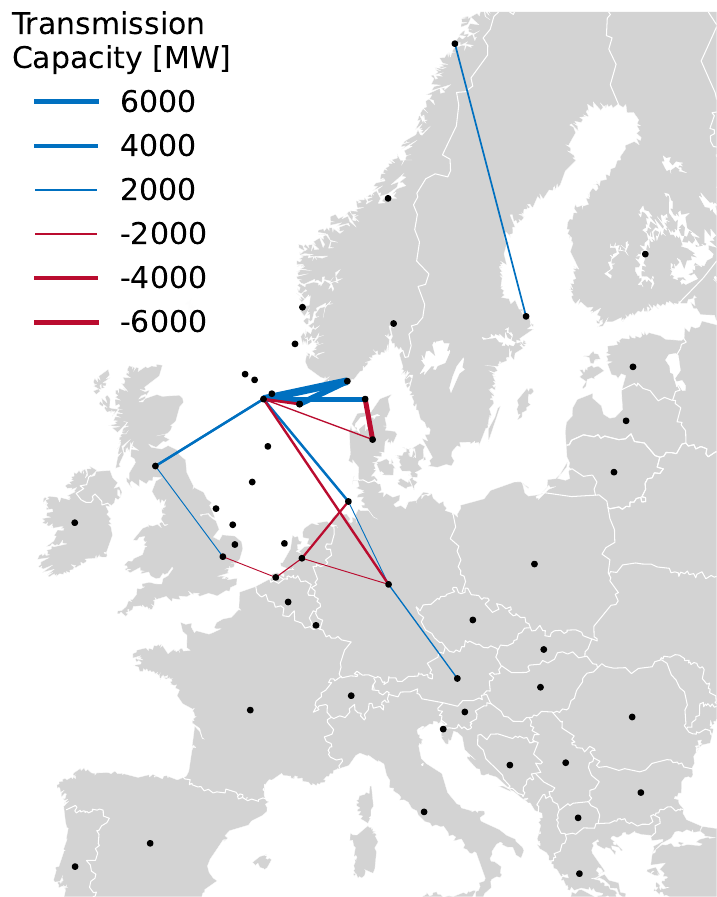}
        \caption{Power transmission capacity difference more than 1000 MW.}
        \label{fig:power_trnsmission_difference}
    \end{subfigure}
    \caption{Power transmission and hydrogen pipeline capacity differences from Base to AST90 in 2045.}
    \label{fig:hydrogen_overview}
\end{figure}

Both hydrogen and power transmission investments increase from Base to AST90, resulting in a total cost rise of €22 billion, as seen in \autoref{fig: OBJ_part1}. Hydrogen production is concentrated in Spain, the United Kingdom, and Norway in AST90, as shown in \autoref{fig: elyzer_pie}. This concentration necessitates more hydrogen transmission pipelines to supply western and central Europe, explaining the increased pipeline costs, as seen in \autoref{fig: pipeline}.

\autoref{fig:power_trnsmission_difference} shows that while total power transmission costs increase, most of the increased capacity will occur in the North Sea, connecting Norway and the United Kingdom to wind farms. Increased hydrogen pipeline capacity in Central and Southwestern Europe will offset the need for additional transmission capacity. More power transmission capacity is developed in mainland Europe between 2024 and 2030 in AST90 to support the rapid increase of \gls{vres}. In 2027-2030, overall power transmission capacity is 10\% higher in AST90 compared to Base. This difference gradually balances out in subsequent years as \gls{vres} capacity differences diminish and higher hydrogen pipeline capacity in AST90 offsets the need for additional power transmission.

\subsection{Breakdown of green hydrogen requirements}

\begin{figure}[!ht]
    \centering
    \includegraphics[width=0.47\textwidth]{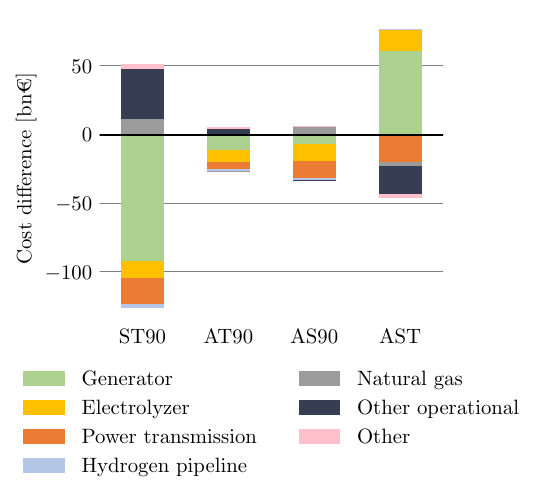}
    \caption{Total system cost differences compared to AST90 in period 2024-2048.}
    \label{fig: cost_difference_AST90}
\end{figure}

\begin{figure*}[t]
    \centering
        \begin{subfigure}[b]{0.3\textwidth}
            \centering
            \includegraphics[width=\textwidth]{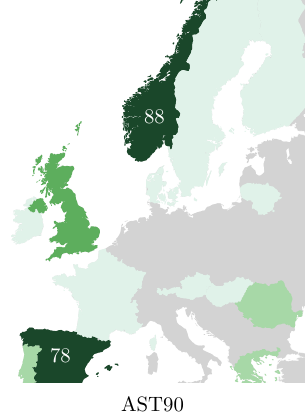}
        \end{subfigure}
        \begin{subfigure}[b]{0.3\textwidth}
            \centering
            \includegraphics[width=\textwidth]{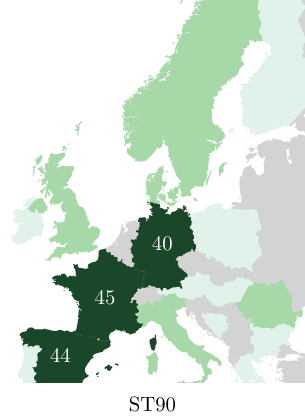}
        \end{subfigure}
        \begin{subfigure}[b]{0.3\textwidth}
            \centering
            \includegraphics[width=\textwidth]{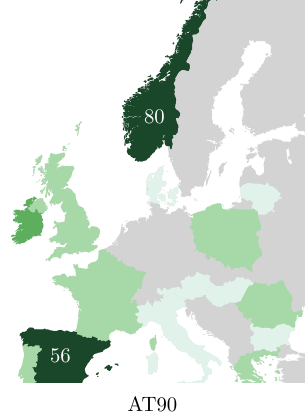}
        \end{subfigure}
        \begin{subfigure}[b]{0.3\textwidth}
            \centering
            \includegraphics[width=\textwidth]{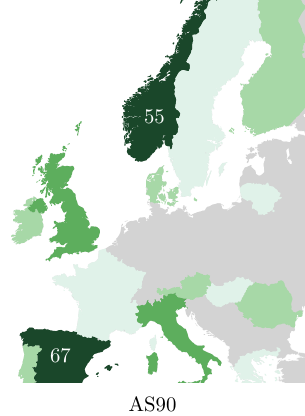}
        \end{subfigure}
        \begin{subfigure}[b]{0.3\textwidth}
            \centering
            \includegraphics[width=\textwidth]{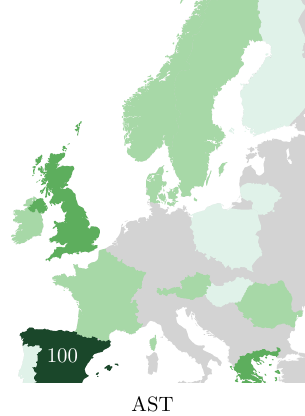}
        \end{subfigure}
        \begin{subfigure}[b]{0.3\textwidth}
            \centering
            \includegraphics[width=\textwidth]{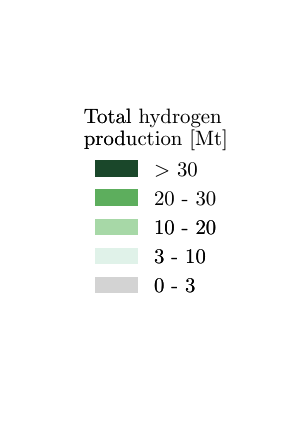}
        \end{subfigure}
    \caption{Total expected green hydrogen production in 2024-2048 per country across all stochastic scenarios.}
    \label{fig:hydrogen_production}
\end{figure*}

\autoref{fig: cost_difference_AST90} shows a breakdown of the cost differences between AST90 and its alternatives where specific requirements are ignored. ST90, which removes the additionality requirement, decreases total system costs by €75 billion, a €7 billion increase from Base. AT90, which removes the spatial correlation requirement, decreases costs by €22 billion, and AS90, which removes the temporal correlation requirement, decreases costs by €28 billion. Conversely, AST increases system costs by €30 billion, as it imposes stricter constraints than AST90 by not exempting some countries. Cost variations are mainly due to differences in investments in electrolyzers and power generators, as well as expected natural gas usage and operational expenditure.

\begin{figure}[h!]
    \centering
    \includegraphics[width=0.47\textwidth]{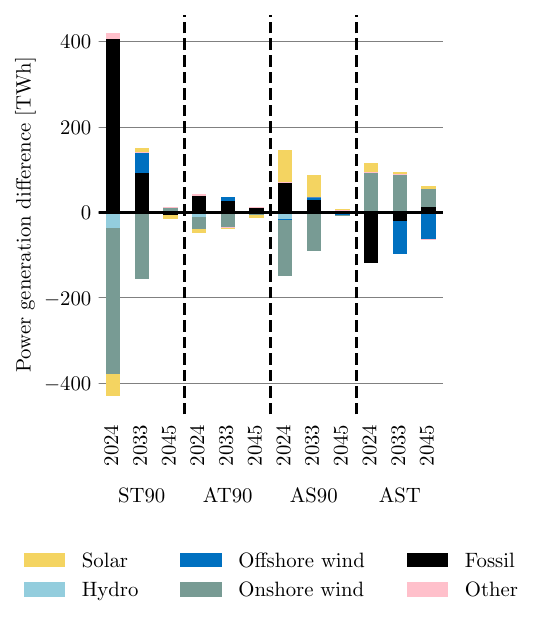}
    \caption{Annual expected power generation differences compared to AST90 in selected years across all stochastic scenarios.}
    \label{fig: gen_diff_part2}
\end{figure}

\autoref{fig:hydrogen_production} displays the total expected hydrogen production in European countries from 2024-2048, and \autoref{fig: gen_diff_part2} shows the expected power generation differences between AST90 and other cases. AST90 highlights dominant hydrogen production in Norway and Spain, each exceeding 78 Mt on expectation. Spain's high production is due to favorable \gls{vres} conditions, particularly onshore wind and solar power. Norway's substantial capacity is attributed to its nearly entirely renewable power grid, exempting it from green hydrogen requirements. Norway's electricity mix is dominated by hydropower and wind.

The results from ST90 suggest that the additionality of \gls{vres} for electrolysis power supply is the costliest green hydrogen requirement. Allowing previously built \gls{vres} capacities to support green hydrogen production shifts new generator investments towards cheaper fossil-based alternatives in early periods. This shift reduces generator and electrolyzer investments, contributing to the total system cost reduction in ST90 compared to AST90. Hydrogen production increases notably in Germany and France, reducing the need for hydrogen pipeline infrastructure due to substantial industrial and transport sector demands. Increased fossil-based power generation indicates that ST90's hydrogen production depends heavily on initial \gls{vres} capacities built for other purposes. Given that the European power market had over 40\% renewables in 2023 \citep{ember2024}, not using this electricity for green hydrogen production becomes expensive.

The AT90 case shows that removing the spatial correlation between \gls{vres} and electrolysis increases hydrogen production in Poland and France compared to AST90. High hydrogen demand in these countries, especially in the industrial and transport sectors, reduces dependency on hydrogen pipelines. AT90 also leads to a decline in total electrolyzer capacity across Europe. Increased spatial flexibility allows electrolyzers to operate at a higher capacity factor, reducing total capacity while maintaining production levels. Power generation differences in AT90 compared to AST90 suggest that more cross-border \gls{vres} exchange helps stabilize electrolyzer production.

Similar results are seen in AS90, where hourly correlation is replaced with annual correlation. Exempt from hourly temporal correlation with \gls{vres}, electrolyzers can partly source power from fossil generators if sufficient renewable power is produced annually. Consequently, some countries achieve higher electrolysis capacity factors. In AS90, hydrogen production is strategically located closer to consumption, including in Italy and Austria. This relocation results in decreased investments in generators, electrolyzers, and hydrogen pipelines compared to AST90. While some countries increase hydrogen production in AS90, they do not surpass those with advantageous \gls{vres} conditions. AS90 also shows increased competitiveness of solar power, as temporal relaxation benefits solar power by matching peak \gls{vres} production during the day and summer with hydrogen production at night and in winter.

Finally, AST results in a geographical shift of hydrogen production from Norway to regions with more cost-efficient potential for new \gls{vres}, particularly Spain. This indicates that Norway's green hydrogen production is economically viable mainly due to the 90\% renewable exemption. Consequently, AST shows decreased fossil-based generation and offshore wind for Norwegian hydrogen production, replaced by increased onshore wind generation elsewhere.

\subsection{Limitations: constraints and assumptions in EMPIRE modeling}

Due to computational limits, data availability, and uncertain future technology development and energy demand, the model has limitations and trade-offs. It does not enforce upper bounds or diseconomies of scale for electrolyzers, potentially underestimating hydrogen production costs. The model assumes rapid \gls{vres} investment is a good strategy for reducing emissions, despite its cost. Scope 3 impacts (indirect lifecycle emissions) are not considered, reflecting their absence in the \gls{euets}, though they have environmental impacts. A model considering total lifetime aspects might reach different conclusions about the transition pace, but this is outside EMPIRE's scope. The 90\% exemption requires a country's grid to have over 90\% renewable power, but ambiguities allow importing non-renewable power without accounting for it in the grid mix.

The model does not consider already planned renewable generator capacity or subsidies, meaning all new \gls{vres} capacity could be eligible for hydrogen production.

\section{Societal and environmental impacts of green hydrogen production}

This analysis aims to calculate the system cost increase of enforcing green hydrogen from electrolyzers using electricity from wave, geothermal, solar, hydropower, and wind. Although the European Commission's definition of renewable hydrogen includes hydrogen production through electrolysis powered by renewable sources (wave, geothermal, solar, hydropower, wind, and nuclear with emissions of maximum 1.2 gCO2eq/MJ) and through biogas reforming or biomass conversion, this study focuses solely on green hydrogen \citep{WorldEconomicForum2021}.

Regulatory constraints on renewable hydrogen production, such as additionality, temporal, and spatial correlation (\autoref{sec:green_def}), have increased system costs by €82 billion. While this paper focuses on the cost increase from implementing green hydrogen from renewable sources, it acknowledges that low-carbon hydrogen, such as hydrogen from fuel combustion, is part of the low-carbon strategy. On September 27, 2024, the European Commission published a draft Delegated Act outlining the methodology for calculating lifecycle emissions to qualify fuels, including low-carbon hydrogen, under Article 9.5 of the Gas Directive. The proposed methodology for achieving a 70\% greenhouse gas emissions reduction includes various hydrogen production methods and emission savings from \gls{ccs} in permanent geological storage and emissions from carbon capture operations \citep{EuropeanCommission2024}.

The €82 billion system cost increase from implementing green hydrogen over 24 years equates to roughly 0.5\% of the projected \gls{eu} GDP for 2025 or 0.4\% of the estimated GDP by 2029 \citep{Statista2024}. While substantial, this cost represents a small portion of broader financial measures, indicating that the investment is not disproportionately high given the \gls{eu}'s economy. Beyond economic impacts, regulatory constraints on green hydrogen can yield societal and environmental benefits. Therefore, the following explores the balance between these regulatory measures' costs and broader contributions.

\subsection{Carbon emission reductions and the role of the emissions trading scheme}

One of the central benefits of green hydrogen production is its potential to reduce carbon emissions within the \gls{euets} framework. The \gls{euets} limits emissions through a cap-and-trade system for high-emission sectors like power generation, industry, and aviation \cite{EuropeanCommission_ETS2024}. This cap decreases over time, encouraging further reductions, and companies can trade emission allowances. Green hydrogen reduces emissions by replacing fossil fuel-based hydrogen production, particularly in hard-to-abate sectors like industry and transport, potentially accelerating decarbonization beyond the \gls{euets}'s direct influence and reducing demand for emission allowances.

However, reduced demand for emission allowances does not necessarily mean reduced emissions within the \gls{euets} due to the politically fixed cap. The waterbed effect \citep{eichner2019eu} suggests that technological innovations within the \gls{euets} mainly lower the cost of meeting climate targets rather than reducing overall emissions. Yet, the assumption of a fixed cap is no longer valid due to allowance cancellation possibilities and evolving \gls{eu} climate targets \citep{perino2018new, waterbed2021bruninx, cifuentes2022european}.

Large-scale green hydrogen deployment involves trade-offs related to climate and environmental impacts. \citet{Shen2024} show that adding 50\% renewable infrastructure for green hydrogen in Europe could reduce lifecycle greenhouse gas emissions by 45\% by 2050, mainly by decarbonizing electricity generation for electrolysis. Green hydrogen reduces emissions by 91\%, compared to 80\% for biofuels and 68\% for hydrogen from fossil fuels with \gls{ccs}. While green hydrogen increases infrastructure emissions, near-zero operational emissions offset this, highlighting long-term carbon benefits despite initial costs.

Although this article does not include a lifecycle assessment, \citet{Shen2024} demonstrate that green hydrogen deployment involves additional trade-offs. Other resource limitations, like water scarcity in Spain, are not modeled. Advanced water management solutions may be needed for hydrogen production in such regions. Future research can explore long-term sustainability and environmental impacts, with all data from this article's analyses available on the public project Github page \citep{empire_green_hydrogen}.

\subsection{Impact of hydrogen demand}

In this analysis, hydrogen production volume remains constant across all scenarios to isolate the cost impacts of changing energy sources for hydrogen production. The model focuses on how energy source changes affect system cost, without considering market-driven demand variations. Higher costs could increase hydrogen prices and reduce demand, while rapid green hydrogen infrastructure expansion might boost demand as derivatives enter the market more quickly. For example, e-fuels for the maritime sector are gaining attention, aligned with initiatives like the FuelEU Maritime regulation and the European Hydrogen Bank auction, where €200 million of the €1.2 billion budget is reserved for maritime sector offtake \citep{InnovationFund2024,EuropeanCommission_FuelEU2024}. This approach allows for a comparison of cost implications across different scenarios, with future studies potentially incorporating dynamic factors such as demand elasticity and government interventions.

\subsection{Energy security as a strategic benefit}

The deployment of green hydrogen can enhance energy security by reducing fossil fuel dependency. Promoting domestic green hydrogen production can help Europe secure a stable and autonomous energy future, as green hydrogen is compatible with renewable energy sources \citep{Lebrouhi2022, hassan2024green}. This transition can shield Europe from external market shocks and aligns with long-term climate and energy resilience objectives. For example, the European Hydrogen Bank’s second auction focuses on projects that avoid dependency on a single third country, which may threaten the supply security of electrolysers. The sourcing of electrolyzer stacks, including surface treatment, cell unit production, and stack assembly from China, should not exceed 25\% \citep{InnovationFund2024}.

\subsection{National hydrogen support policies and production locations}

This article covers 52 nodes (\autoref{sec:data}) and takes a pan-European approach, excluding national policies. However, national policies significantly influence hydrogen production in Europe. For instance, the European Commission approved €998 million in state aid for the Netherlands in July 2024, and Germany plans a 9,700-kilometer hydrogen pipeline, pending an EU state aid decision \citep{NLTimes2024,CleanEnergyWire2024a,CleanEnergyWire2024b}. The German H2Global initiative, with €900 million in initial funding and €3.5 billion for future auctions, aims to boost international green hydrogen trade \citep{NLTimes2024}. Increased financial and policy support could accelerate hydrogen projects and influence Europe's hydrogen landscape.

Despite regulatory and cost challenges, green hydrogen offers long-term environmental and societal benefits, reducing carbon emissions and enhancing energy security, making it vital for Europe’s sustainable energy future.

\section{Conclusion and policy implications} \label{ch:conclusion}
Using a stochastic capacity expansion model of the European energy system, this paper estimates the impact of ensuring all hydrogen production in Europe is green, per the \gls{eu} definition \citep{euRenewableHydrogen2023}. The cost of labeling all European hydrogen green was determined by comparing a scenario with all green hydrogen policies enforced to one with no rules. Additionally, a second case study examined investment differences when relaxing green hydrogen policies. All scenarios assume hydrogen production exclusively with electrolyzers.

It is important to note that ensuring all future hydrogen production in Europe is green is neither feasible in the short-term nor the only pathway compatible with the EU strategy. Several simplifications to the \gls{eu} green hydrogen definition were made to obtain quantitative results, including ignoring exemptions to the additionality rule and assuming stricter hourly temporal correlation without a phased-in compliance period. Therefore, the results are upper bound cost estimates with their respective techno-economic implications.

The net additional cost of exclusively permitting green hydrogen in Europe is estimated at €82 billion over 24 years. This upper estimate compares a scenario with green hydrogen policies to one where hydrogen is produced solely using electrolyzers not necessarily sources from new renewable energy sources. Most cost increases occur in new \gls{vres} investments, mainly onshore wind, followed by additional electrolyzer and energy transmission investments. More electrolyzers are needed to adapt to \gls{vres} availability for green hydrogen. Increased renewable power demand accelerates the phase-out of fossil generators, reducing fossil power production by an average of 108 TWh annually between 2024 and 2030. 

The impact of removing each requirement or exemption in the criteria for green hydrogen is as follows:

\begin{itemize}
    \item [] \textbf{Additionality:} Removing additionality requirements significantly reduces total system costs by lowering the need for generator and electrolyzer investments. This leads to more fossil-based power production, especially before 2030. Allowing green hydrogen production from existing \gls{vres} capacities could greatly lower costs.
    
    \item [] \textbf{Spatial correlation:} Removing spatial correlation reduces costs as fewer investments in generators and electrolyzers are needed, allowing them to be built in more favorable locations. Hydrogen production is reallocated closer to high-demand areas, using imported \gls{vres}, resulting in the same production quantities from lower electrolysis capacities.
   
    \item [] \textbf{Temporal correlation:} Changing hourly matching to annual matching reduces total system costs. This leads to more evenly distributed hydrogen production in Europe, with countries having less favorable \gls{vres} conditions becoming more competitive. Solar power also becomes more competitive as a source for green hydrogen production.
    
    \item [] \textbf{90\% exemption:} Removing the 90\% renewable grid exemption increases total system costs. Norwegian hydrogen production is replaced by increased production in Sweden, Spain, and France. Reduced hydrogen production in Norway and lower fossil-based power generation suggest that the renewable grid exemption can significantly lower system costs for green hydrogen policies.
\end{itemize}

This analysis shows that, despite significant net additional cost estimates under stricter rules than those suggested by the \gls{eu}, these costs represent a relatively small percentage of the projected \gls{eu} GDP. This highlights the economic feasibility of such investments for long-term sustainability. Green hydrogen’s potential to reduce \gls{co2} emissions in hard-to-abate sectors further supports these investments as part of Europe's decarbonization strategy.

Currently, EMPIRE limits operational hours to representative weeks per season, preventing stored energy in hydrogen caverns, gas tanks, or pumped hydro stations from being transferred between seasons. Future research should consider long-term hydrogen storage in stochastic optimization models to better understand its potential to balance energy markets across seasonal variations. Additionally, more research is needed to refine the implementation of green hydrogen rules, including all exceptions, as they evolve within the \gls{eu}.

\printcredits

\section*{Declaration of competing interest}
{The authors declare that they have no known competing financial interests or personal relationships that could have appeared to influence the work reported in this paper.}

\section*{Acknowledgements}
This publication has been partially funded by the Hydrogen Pathways 2050 project - Transition of the Norwegian society and value creation from export (project code 326769) and the NordicH2ubs project - Roadmaps towards 2030 and 2040 (project code 346870). The authors gratefully acknowledge the financial support from the Research Council of Norway and the projects' user partners.

\section*{Declaration of Generative AI and AI-assisted technologies in the writing process}
During the preparation of this work, ChatGPT and Microsoft Copilot has been used for language polishing. After using this tool/service, the authors reviewed and edited the content as needed and takes full responsibility for the content of the publication.



\newglossaryentry{balmorel}{
    name={Balmorel},
    description={Baltic Model of Regional Electricity Liberalization}
}
\newglossaryentry{empire}{
    name={EMPIRE},
    description={European Model for Power System Investment with high shares of Renewable Energy}
}
\newglossaryentry{miret-eu}{
    name={MIRET-EU},
    description={Model of International Renewable Energy Transmission - European Union}
}
\newglossaryentry{pw}{
    name={PW},
    description={Power}
}
\newglossaryentry{pypsa-eur}{
    name={PyPSA-Eur},
    description={Python for Power System Analysis - Europe}
}
\newglossaryentry{pypsa-eur-sec}{
    name={PyPSA-Eur-Sec},
    description={Python for Power System Analysis - Europe - Sector Coupling}
}
\newglossaryentry{red2}{
    name={RED II},
    description={Renewable Energy Directive II}
}
\newglossaryentry{gdp}{
    name={GDP},
    description={Gross domestic product}
}

\printglossary[type=\acronymtype, title=Acronyms]

\section*{Glossary}

\begin{longtable}{p{3cm}p{12cm}}
\textbf{Term} & \textbf{Description} \\ \hline
Balmorel & Baltic Model of Regional Electricity Liberalization \\
EMPIRE & European Model for Power System Investment with high shares of Renewable Energy \\
MIRET-EU & Model of International Renewable Energy Transmission - European Union \\
PW & Power \\
PyPSA-Eur & Python for Power System Analysis - Europe \\
PyPSA-Eur-Sec & Python for Power System Analysis - Europe - Sector Coupling \\
RED II & Renewable Energy Directive II \\
\end{longtable}

\clearpage
\bibliographystyle{cas-model2-names}

\bibliography{lit}

\newpage
\onecolumn


\appendix
\renewcommand{\thesection}{\Alph{section}} 
\newcommand{\sectionname}{Appendix}

\section{Mathematical formulation}
\label{appendix_a}

\begin{mdframed} 
\setlist[description]{style=standard,leftmargin=2cm,labelindent=0cm,labelwidth=1.5cm,labelsep=0.5cm,align=left} 
\begin{description}
    \item[$\omega$] Operational scenario
    \item[$c$] Commodity
    \item[$h$] Operational hour
    \item[$i,j$] Investment period
    \item[$n,m$] Node
    \item[$p$] Production method
    \item[$s$] Season
    \item[$b$] Storage type
    \item[$t$] Transmission type

    \item[$\mathcal{C}$] Commodities
    \item[$\mathcal{N}$] Nodes
    \item[$\mathcal{H}$] Operational hours
    \item[$\mathcal{H}^F$] First hour of every season
    \item[$\mathcal{H}^L$] Last hour of every season
    \item[$\mathcal{H}^s$] Hours belonging to season $s$
    \item[$\mathcal{I}$] Investment periods
    \item[$\mathcal{L}^c_n$] All possible bidirectional arcs to node $n$ for commodity $c$
    \item[$\mathcal{P}^c$] Production methods for commodity $c$
    \item[$\mathcal{B}^c$] Storage types for commodity $c$
    \item[$\mathcal{T}^c$] Transmission types for commodity $c$
    \item[$\mathcal{S}$] Seasons
    \item[$\Omega$] Operational Scenarios
    \item[$\sigma^c$] Sinks of commodity $c$

    \item[$D^c_{n,h,i,\omega}$] Exogenous demand for commodity $c$ in node $n$, hour $h$, period $i$, scenario $\omega$
    \item[$A^c_n$] Total capacity for commodity $c$ in node $n$
    \item[$i^{life}_p$] Lifetime of production method $p$
    \item[$i^{life}_{t}$] Lifetime of transmission type $t$
    \item[$i^{life}_{b}$] Lifetime of storage type $b$
    \item[$L^{period}$] Length of investment periods
    \item[$\bar x^p_{n,i}$] Remaining initial capacity of production method $p$ in node $n$, period $i$
    \item[$\bar x^b_{n,i}$] Remaining initial capacity of storage type $b$ in node $n$, period $i$
    \item[$\bar x^t_{n,m,i}$] Remaining initial capacity of transmission type $t$ for bidirectional arc $(n,m)$, in period $i$
    \item[$\alpha_s$] Scale factor for season $s$
    \item[$\pi_\omega$] Probability of scenario $\omega$
    \item[$r$] Annual discount rate

    \item[$v^p_{n,i}$] Available capacity of production method $p$ in node $n$, period $i$
    \item[$v^b_{n,i}$] Available capacity of storage type $b$ in node $n$, period $i$
    \item[$v^t_{n,m,i}$] Available capacity of transmission type $t$ in bidirectional arc $(n,m)$, period $i$
    \item[$x^p_{n,i}$] Capacity built of production method $p$ in node $n$, period $i$
    \item[$x^b_{n,i}$] Capacity built of storage type $b$ in node $n$, period $i$
    \item[$x^t_{n,m,i}$] Capacity built of transmission type $t$ in bidirectional arc $(n,m)$, period $i$
    \item[$y^{c,trans}_{t,n,m,h,i,\omega}$] Transmission at transmission type $t$ for commodity $c$ in bidirectional arc $(n,m)$, hour $h$, period $i$, scenario $\omega$
    \item[$y^{c,sink}_{n,h,i,\omega}$] Endogenous demand of commodity $c$ in node $n$, hour $h$, period $i$, scenario $\omega$
    \item[$y^{c,source}_{p,n,h,i,\omega}$] Production of commodity $c$ by production method $p$ in node $n$, hour $h$, period $i$, scenario $\omega$
    \item[$y^{c,ll}_{n,h,i,\omega}$] Load shed of commodity $c$ in node $n$, hour $h$, period $i$, scenario $\omega$
    \item[$y^{c,chrg}_{b,n,h,i,\omega}$] Charging of storage type $b$ for commodity $c$ in node $n$, hour $h$, period $i$, scenario $\omega$
    \item[$y^{c,dischrg}_{b,n,h,i,\omega}$] Discharging of storage type $b$ for commodity $c$ in node $n$, hour $h$, period $i$, scenario $\omega$
    \item[$y^{c,stor}_{b,n,h,i,\omega}$] Storage level of commodity $c$ in storage type $b$, node $n$, hour $h$, period $i$, scenario $\omega$

    \item[$g$] Power generator
    \item[$e$] Electrolyzer

    \item[$\mathcal{G}$] Power generators
    \item[$\mathcal{G}^{VRES}$] Power generators defined as \gls{vres}
    \item[$\mathcal{E}$] Electrolyzers
    \item[$\mathcal{N}^{90}_i$] Nodes that qualify for 90\% renewable grid exemption in period $i$
    \item[$\eta^{PW}_e$] Constant power consumption for producing one ton hydrogen at electrolyzer $e$
    \item[$\alpha_{g,n,h,i,\omega}$] Availability of variable power generator $g$ in node $n$, hour $h$, period $i$, scenario $\omega$
    \item[$H_i$] Fixed total hydrogen production quantity from electrolysis in period $i$

    \item[$\kappa^{repurpose}$] Repurpose cost factor
    \item[$\eta^{repurpose}$] Repurpose energy flow factor
    \item[$x^{{repurpose}}_{n,m,i}$] Lost natural gas pipeline capacity in bidirectional arc $(n,m)$ in period $i$ due to repurposing
\end{description}
\end{mdframed}

As mentioned in \autoref{ch:method}, EMPIRE is a multi-horizon, stochastic energy system model that minimizes total system costs to meet energy demand in Europe. These costs include investments in production, storage, and transmission, as well as operational and commodity load shed costs. All costs are discounted at an annual rate $r$ of 5\%, where $v = \sum_{j=0}^{(L^{period}-1)} (1+r)^{-j}$ represents the scaling and discounting of annualized costs over the investment period. The objective function is described in \autoref{eq:objective}.

\small
\begin{align}\label{eq:objective}
    \begin{split}
        \min z = \sum_{i \in \mathcal{i}} (1+r)^{L^{period}(i-1)} \times \\
        [\sum_{c \in \mathcal{C}} (\sum_{n \in \mathcal{N}}\sum_{a \in \mathcal{P}^c \cup \mathcal{B}^c} q^{inv}_{a,i} x^a_{n,i} + \sum_{n \in \mathcal{N}} \sum_{m \in \mathcal{L}^c_n} \sum_{t \in \mathcal{T}^c} q^{inv}_{t,i} x^t_{n,m,i}) + \\
        v \sum_{\omega \in \Omega} \pi_\omega \sum_{s \in \mathcal{S}} \alpha_s \sum_{h \in \mathcal{H}^s}\sum_{n \in \mathcal{N}}\sum_{c \in \mathcal{C}}\sum_{p \in \mathcal{P}^c} q^{operational}_{p,i} y^{c,source}_{p,n,h,i,\omega} + \\
        v \sum_{\omega \in \Omega} \pi_\omega \sum_{s \in \mathcal{S}} \alpha_s \sum_{h \in \mathcal{H}^s}\sum_{n \in \mathcal{N}}\sum_{c \in \mathcal{C}} q^{c,ll}_{n,i} y^{c,ll}_{n,h,i,\omega}].
    \end{split}
\end{align}
\normalsize

A general formulation of the flow balance is used for a commodity $c$ in EMPIRE, covering power, hydrogen, natural gas, \gls{ccs} transport, steel, ammonia, cement, and refinery sectors. \autoref{eq:balance} states that the demand for a commodity $D^c_{n,h,i,\omega}$ must be balanced by the sum of production, endogenous use, net storage charge, net export, and load shed.

\small
\begin{align}\label{eq:balance}
\begin{split}
\sum_{p \in \mathcal{P}^c} y^{c,source}_{p,n,h,i,\omega} - \sum_{sink \in \sigma} y^{c,sink}_{n,h,i,\omega} - \sum_{\mathcal{B}^c}(y^{c,chrg}_{b,n,h,i,\omega} - y^{c,dischrg}_{b,n,h,i,\omega}) -\\
\sum_{\mathcal{T}^c}\sum_{m \in \mathcal{L}_{n}^{c}}( y^{c,trans}_{t,n,m,h,i,\omega} - y^{c,trans}_{t,m,n,h,i,\omega} ) + y^{c,ll}_{n,h,i,\omega} = D^c_{n,h,i,\omega} \\ \forall ~c \in \mathcal{C},~n \in \mathcal{N}, ~h \in \mathcal{H}, ~i \in \mathcal{I}, ~\omega \in \Omega.
\end{split}
\end{align}
\normalsize

\autoref{eq:lifetime_production}, \autoref{eq:lifetime_storage}, and \autoref{eq:lifetime_transmission} define the total available capacity of production $v_{n,i}^{p}$, storage $v_{n,i}^{b}$, and transmission $v_{n,m,i}^{t}$ as the sum of all invested capacity within its lifetime. $i^{'}$ represents the first investment period still within the asset's lifetime, relative to the current period $i$. Total available capacity equals the sum of the invested and initial capacity.

\small
\begin{align}\label{eq:lifetime_production}
\begin{split}
\sum_{j=i'}^{i}x_{n,j}^p + \bar x_{n,i}^p=v_{n,i}^p \\
\forall ~c \in \mathcal{C},~p \in \mathcal{P}^c,~n \in \mathcal{N}, 
~i \in \mathcal{I},~i^{'}=max\{1,i-i_p^{life}\}.    
\end{split}
\end{align}

\begin{align}\label{eq:lifetime_storage}
\begin{split}
\sum_{j=i'}^{i}x_{n,j}^b + \bar x_{n,i}^b=v_{n,i}^b \\
\forall ~c \in \mathcal{C},~b \in \mathcal{B}^c,~n \in \mathcal{N}, 
~i \in \mathcal{I},~i^{'}=max\{1,i-i_b^{life}\}.    
\end{split}
\end{align}

\begin{align}\label{eq:lifetime_transmission}
\begin{split}
\sum_{j=i'}^{i}x_{n,m,j}^{t} + \bar x_{n,m,i}^{t}=v_{n,m,i}^{t} \\
\forall ~c \in \mathcal{C},~t \in \mathcal{T}^c,~n \in \mathcal{N}, ~m \in \mathcal{L}^c_n, \\
~i \in \mathcal{I},~i^{'}=max\{1,i-i_t^{life}\}.   
\end{split}
\end{align}
\normalsize

\autoref{eq:max_operation_production}, \autoref{eq:max_operation_storage} and \autoref{eq:max_operation_transmisssion} assures that assets cannot be operated above the installed capacity. 

\small
\begin{align}\label{eq:max_operation_production}
    \begin{split}
        y^{c,source}_{p,n,h,i,\omega} \leq v_{n,i}^{p} \quad \\
        \forall ~c \in \mathcal{C},~p \in \mathcal{P}^c,~n \in \mathcal{N},
        ~h \in \mathcal{H},
        ~i \in \mathcal{I}, ~\omega \in \Omega. 
    \end{split}
\end{align}

\begin{align}\label{eq:max_operation_storage}
    \begin{split}
        y^{c,stor}_{b,n,h,i,\omega} \leq v_{n,i}^{b} \quad \\
        \forall ~c \in \mathcal{C},~b \in \mathcal{B}^c,~n \in \mathcal{N}, 
        ~h \in \mathcal{H},
        ~i \in \mathcal{I}, ~\omega \in \Omega. 
    \end{split}
\end{align}

\begin{align}\label{eq:max_operation_transmisssion}
\begin{split}
    y^{c,trans}_{t,n,m,h,i,\omega} \leq v_{n,i}^{t} \quad \\
    \forall ~c \in \mathcal{C},~t \in \mathcal{T}^c,~n \in \mathcal{N},~m \in \mathcal{L}^c_n, \\
    ~h \in \mathcal{H},
    ~i \in \mathcal{I}, ~\omega \in \Omega.
\end{split}
\end{align}
\normalsize

\autoref{eq:storage_level} states that the current storage level equals the previous level plus the net charge.

\small
\begin{align}\label{eq:storage_level}
\begin{split}
    y^{c,stor}_{b,n,h-1,i,\omega} + y^{c,chrg}_{b,n,h,i,\omega} - y^{c,dischrg}_{b,n,h,i,\omega} = y^{c,stor}_{b,n,h,i,\omega}\\
    \forall ~c \in \mathcal{C},~b \in \mathcal{B}^c,~n \in \mathcal{N}, ~h \in \mathcal{H} \setminus \mathcal{H}^{F},~i \in \mathcal{I},~\omega \in \Omega.    
\end{split}
\end{align}
\normalsize

Storage level at the beginning and end of a season equals half the installed capacity of storage to remain in energy balance, as described in \autoref{eq:storage_start} and \autoref{eq:storage_end} respectively. No storage is transferred between seasons.

\small
\begin{align}\label{eq:storage_start}
\begin{split}
0.5 \times v^{b}_{n,i} + y^{c,chrg}_{b,n,h,i,\omega} - y^{c,dischrg}_{b,n,h,i,\omega} = y^{c,stor}_{b,n,h,i,\omega}\\
\forall ~c \in \mathcal{C},~b \in \mathcal{B}^c,~n \in \mathcal{N}, ~h \in \mathcal{H}^{F}, ~i \in \mathcal{I}, ~\omega \in \Omega.
\end{split}
\end{align}

\begin{align}\label{eq:storage_end}
\begin{split}
0.5 \times v^{b}_{n,i} = y^{c,stor}_{b,n,h,i,\omega} \\
\forall ~c \in \mathcal{C},~b \in \mathcal{B}^c,~n \in \mathcal{N}, ~h \in \mathcal{H}^L, ~i \in \mathcal{I}, ~\omega \in \Omega.  
\end{split}
\end{align}

\normalsize

Some resources have absolute limits over all periods. \autoref{eq:limit_source} and \autoref{eq:limit_sink} address these limits and consider the limit of natural gas production and \gls{co2} sequestering, respectively. The total production $y^{c,source}_{p,n,h,i,\omega}$ or endogenous use $y^{c,sink}_{n,h,i,\omega}$, scaled by seasonal scale $\alpha_s$ and length of investment periods $L^{period}$, is limited by total available capacity $A^c_n$.

\small
\begin{align}
\label{eq:limit_source}
\begin{split}
\sum_{i \in \mathcal{I}}\sum_{s \in \mathcal{S}} \sum_{h \in \mathcal{H}^s}\sum_{p \in \mathcal{P}^c} L^{period} \times \alpha_s \times y^{c,source}_{p,n,h,i,\omega} \leq A^{c}_{n} \\
\forall ~c \in \{NG\},~n \in \mathcal{N}, ~\omega \in \Omega.
\end{split}
\end{align}

\begin{align}
\label{eq:limit_sink}
\begin{split}
\sum_{i \in \mathcal{I}}\sum_{s \in \mathcal{S}} \sum_{h \in \mathcal{H}^s} L^{period} \times \alpha_s \times y^{c,sink}_{n,h,i,\omega} \leq A^{c}_{n} \\
\forall ~c \in \{\text{\gls{co2}}\}~n \in \mathcal{N}, ~\omega \in \Omega.
\end{split}
\end{align}
\normalsize

Finally, \autoref{eq:emission_cap} ensures that the total emission in each scenario is limited by the emission cap $E^{max}_i$ for each investment period.

\small
\begin{align}\label{eq:emission_cap}
    \begin{split}
        \sum_{s \in \mathcal{S}} \alpha_s \sum_{h \in \mathcal{H}^s}\sum_{n \in \mathcal{N}}\sum_{c \in \mathcal{C}}\sum_{p \in \mathcal{P}^c} \eta_p^{emission} y^{c,source}_{p,n,h,i,\omega} \leq E^{max}_i\quad \\
        \forall ~i \in \mathcal{I},~\omega \in \Omega. 
    \end{split}
\end{align}
\normalsize

\clearpage

\section{Renewable grid estimates}
\label{appendix_b}
Table \ref{tab:90percentshare_vres} shows whether countries in EMPIRE are exempt from green hydrogen rules due to having a national grid with over 90\% renewable share, as defined by \cite{euRenewableHydrogen2023}. This input makes the model computationally tractable. Ideally, a country's renewable share would be an endogenous decision based on existing and future capacities when simulating the cost-optimal development of the European electricity market in line with climate policies. However, since this decision affects rule exemptions, it is assumed ex-ante based on \citet{durakovic2024decarbonizing}.

\autoref{eq:renewable_grid_exemption} ensures that exempt countries in Table \ref{tab:90percentshare_vres} maintain a renewable share of 90\% or more within the investment period. While this imposes political constraints, it is necessary for modeling the 90\% rule in a computationally feasible way using a European multi-horizon stochastic programming model.

Although many countries in Table \ref{tab:90percentshare_vres} are not assumed to be exempt from green hydrogen rules, they may still decide their renewable share towards 2048, and any country can achieve a renewable share of 90\% or more.

\clearpage

\begin{table}[ht]
\centering
\scriptsize
\caption{Input on when and if a country must maintain renewable energy production shares above 90\% in its national grid from 2024 to 2048 based on results from the EMPIRE model version presented in \citet{durakovic2024decarbonizing}.} \label{tab:90percentshare_vres}
\begin{tabular}{l|llllllll}
\cellcolor[HTML]{FFFFFF}                   & \cellcolor[HTML]{FFFFFF}                                    & \cellcolor[HTML]{FFFFFF}                                    & \cellcolor[HTML]{FFFFFF}                                    & \cellcolor[HTML]{FFFFFF}                                    & \cellcolor[HTML]{FFFFFF}                                    & \cellcolor[HTML]{FFFFFF}                                    & \cellcolor[HTML]{FFFFFF}                                    & \cellcolor[HTML]{FFFFFF}                                    \\
\multirow{-2}{*}{\cellcolor[HTML]{FFFFFF}} & \multirow{-2}{*}{\cellcolor[HTML]{FFFFFF}2024-2027}         & \multirow{-2}{*}{\cellcolor[HTML]{FFFFFF}2027-2030}         & \multirow{-2}{*}{\cellcolor[HTML]{FFFFFF}2030-2033}         & \multirow{-2}{*}{\cellcolor[HTML]{FFFFFF}2033-2036}         & \multirow{-2}{*}{\cellcolor[HTML]{FFFFFF}2036-2039}         & \multirow{-2}{*}{\cellcolor[HTML]{FFFFFF}2039-2042}         & \multirow{-2}{*}{\cellcolor[HTML]{FFFFFF}2042-2045}         & \multirow{-2}{*}{\cellcolor[HTML]{FFFFFF}2045-2048}         \\ \hline
                                           &                                                             &                                                             &                                                             &                                                             &                                                             & \cellcolor[HTML]{C6E0B4}                                    & \cellcolor[HTML]{C6E0B4}                                    & \cellcolor[HTML]{C6E0B4}                                    \\
\multirow{-2}{*}{Austria}                  & \multirow{-2}{*}{-}                                         & \multirow{-2}{*}{-}                                         & \multirow{-2}{*}{-}                                         & \multirow{-2}{*}{-}                                         & \multirow{-2}{*}{-}                                         & \multirow{-2}{*}{\cellcolor[HTML]{C6E0B4}\textgreater 90\%} & \multirow{-2}{*}{\cellcolor[HTML]{C6E0B4}\textgreater 90\%} & \multirow{-2}{*}{\cellcolor[HTML]{C6E0B4}\textgreater 90\%} \\
                                           &                                                             &                                                             &                                                             &                                                             &                                                             &                                                             &                                                             &                                                             \\
\multirow{-2}{*}{Belgium}                  & \multirow{-2}{*}{-}                                         & \multirow{-2}{*}{-}                                         & \multirow{-2}{*}{-}                                         & \multirow{-2}{*}{-}                                         & \multirow{-2}{*}{-}                                         & \multirow{-2}{*}{-}                                         & \multirow{-2}{*}{-}                                         & \multirow{-2}{*}{-}                                         \\
                                           &                                                             &                                                             &                                                             &                                                             &                                                             &                                                             &                                                             &                                                             \\
\multirow{-2}{*}{BosniaH}                  & \multirow{-2}{*}{-}                                         & \multirow{-2}{*}{-}                                         & \multirow{-2}{*}{-}                                         & \multirow{-2}{*}{-}                                         & \multirow{-2}{*}{-}                                         & \multirow{-2}{*}{-}                                         & \multirow{-2}{*}{-}                                         & \multirow{-2}{*}{-}                                         \\
                                           &                                                             &                                                             &                                                             &                                                             &                                                             &                                                             & \cellcolor[HTML]{C6E0B4}                                    & \cellcolor[HTML]{C6E0B4}                                    \\
\multirow{-2}{*}{Bulgaria}                 & \multirow{-2}{*}{-}                                         & \multirow{-2}{*}{-}                                         & \multirow{-2}{*}{-}                                         & \multirow{-2}{*}{-}                                         & \multirow{-2}{*}{-}                                         & \multirow{-2}{*}{-}                                         & \multirow{-2}{*}{\cellcolor[HTML]{C6E0B4}\textgreater 90\%} & \multirow{-2}{*}{\cellcolor[HTML]{C6E0B4}\textgreater 90\%} \\
                                           &                                                             &                                                             &                                                             &                                                             &                                                             &                                                             &                                                             &                                                             \\
\multirow{-2}{*}{Croatia}                  & \multirow{-2}{*}{-}                                         & \multirow{-2}{*}{-}                                         & \multirow{-2}{*}{-}                                         & \multirow{-2}{*}{-}                                         & \multirow{-2}{*}{-}                                         & \multirow{-2}{*}{-}                                         & \multirow{-2}{*}{-}                                         & \multirow{-2}{*}{-}                                         \\
                                           &                                                             &                                                             &                                                             &                                                             &                                                             &                                                             &                                                             &                                                             \\
\multirow{-2}{*}{CzechR}                   & \multirow{-2}{*}{-}                                         & \multirow{-2}{*}{-}                                         & \multirow{-2}{*}{-}                                         & \multirow{-2}{*}{-}                                         & \multirow{-2}{*}{-}                                         & \multirow{-2}{*}{-}                                         & \multirow{-2}{*}{-}                                         & \multirow{-2}{*}{-}                                         \\
                                           &                                                             &                                                             &                                                             &                                                             & \cellcolor[HTML]{C6E0B4}                                    & \cellcolor[HTML]{C6E0B4}                                    & \cellcolor[HTML]{C6E0B4}                                    & \cellcolor[HTML]{C6E0B4}                                    \\
\multirow{-2}{*}{Denmark}                  & \multirow{-2}{*}{-}                                         & \multirow{-2}{*}{-}                                         & \multirow{-2}{*}{-}                                         & \multirow{-2}{*}{-}                                         & \multirow{-2}{*}{\cellcolor[HTML]{C6E0B4}\textgreater 90\%} & \multirow{-2}{*}{\cellcolor[HTML]{C6E0B4}\textgreater 90\%} & \multirow{-2}{*}{\cellcolor[HTML]{C6E0B4}\textgreater 90\%} & \multirow{-2}{*}{\cellcolor[HTML]{C6E0B4}\textgreater 90\%} \\
                                           &                                                             &                                                             &                                                             &                                                             &                                                             &                                                             &                                                             &                                                             \\
\multirow{-2}{*}{Estonia}                  & \multirow{-2}{*}{-}                                         & \multirow{-2}{*}{-}                                         & \multirow{-2}{*}{-}                                         & \multirow{-2}{*}{-}                                         & \multirow{-2}{*}{-}                                         & \multirow{-2}{*}{-}                                         & \multirow{-2}{*}{-}                                         & \multirow{-2}{*}{-}                                         \\
                                           &                                                             &                                                             &                                                             &                                                             &                                                             &                                                             &                                                             &                                                             \\
\multirow{-2}{*}{Finland}                  & \multirow{-2}{*}{-}                                         & \multirow{-2}{*}{-}                                         & \multirow{-2}{*}{-}                                         & \multirow{-2}{*}{-}                                         & \multirow{-2}{*}{-}                                         & \multirow{-2}{*}{-}                                         & \multirow{-2}{*}{-}                                         & \multirow{-2}{*}{-}                                         \\
                                           &                                                             &                                                             &                                                             &                                                             &                                                             &                                                             & \cellcolor[HTML]{C6E0B4}                                    & \cellcolor[HTML]{C6E0B4}                                    \\
\multirow{-2}{*}{France}                   & \multirow{-2}{*}{-}                                         & \multirow{-2}{*}{-}                                         & \multirow{-2}{*}{-}                                         & \multirow{-2}{*}{-}                                         & \multirow{-2}{*}{-}                                         & \multirow{-2}{*}{-}                                         & \multirow{-2}{*}{\cellcolor[HTML]{C6E0B4}\textgreater 90\%} & \multirow{-2}{*}{\cellcolor[HTML]{C6E0B4}\textgreater 90\%} \\
                                           &                                                             &                                                             &                                                             &                                                             &                                                             &                                                             &                                                             &                                                             \\
\multirow{-2}{*}{Germany}                  & \multirow{-2}{*}{-}                                         & \multirow{-2}{*}{-}                                         & \multirow{-2}{*}{-}                                         & \multirow{-2}{*}{-}                                         & \multirow{-2}{*}{-}                                         & \multirow{-2}{*}{-}                                         & \multirow{-2}{*}{-}                                         & \multirow{-2}{*}{-}                                         \\
                                           &                                                             &                                                             &                                                             &                                                             &                                                             &                                                             &                                                             & \cellcolor[HTML]{C6E0B4}                                    \\
\multirow{-2}{*}{GreatBrit.}               & \multirow{-2}{*}{-}                                         & \multirow{-2}{*}{-}                                         & \multirow{-2}{*}{-}                                         & \multirow{-2}{*}{-}                                         & \multirow{-2}{*}{-}                                         & \multirow{-2}{*}{-}                                         & \multirow{-2}{*}{-}                                         & \multirow{-2}{*}{\cellcolor[HTML]{C6E0B4}\textgreater 90\%} \\
                                           &                                                             &                                                             &                                                             &                                                             & \cellcolor[HTML]{C6E0B4}                                    & \cellcolor[HTML]{C6E0B4}                                    & \cellcolor[HTML]{C6E0B4}                                    & \cellcolor[HTML]{C6E0B4}                                    \\
\multirow{-2}{*}{Greece}                   & \multirow{-2}{*}{-}                                         & \multirow{-2}{*}{-}                                         & \multirow{-2}{*}{-}                                         & \multirow{-2}{*}{-}                                         & \multirow{-2}{*}{\cellcolor[HTML]{C6E0B4}\textgreater 90\%} & \multirow{-2}{*}{\cellcolor[HTML]{C6E0B4}\textgreater 90\%} & \multirow{-2}{*}{\cellcolor[HTML]{C6E0B4}\textgreater 90\%} & \multirow{-2}{*}{\cellcolor[HTML]{C6E0B4}\textgreater 90\%} \\
                                           &                                                             &                                                             &                                                             &                                                             &                                                             &                                                             & \cellcolor[HTML]{C6E0B4}                                    & \cellcolor[HTML]{C6E0B4}                                    \\
\multirow{-2}{*}{Hungary}                  & \multirow{-2}{*}{-}                                         & \multirow{-2}{*}{-}                                         & \multirow{-2}{*}{-}                                         & \multirow{-2}{*}{-}                                         & \multirow{-2}{*}{-}                                         & \multirow{-2}{*}{-}                                         & \multirow{-2}{*}{\cellcolor[HTML]{C6E0B4}\textgreater 90\%} & \multirow{-2}{*}{\cellcolor[HTML]{C6E0B4}\textgreater 90\%} \\
                                           &                                                             &                                                             &                                                             &                                                             &                                                             &                                                             &                                                             &                                                             \\
\multirow{-2}{*}{Ireland}                  & \multirow{-2}{*}{-}                                         & \multirow{-2}{*}{-}                                         & \multirow{-2}{*}{-}                                         & \multirow{-2}{*}{-}                                         & \multirow{-2}{*}{-}                                         & \multirow{-2}{*}{-}                                         & \multirow{-2}{*}{-}                                         & \multirow{-2}{*}{-}                                         \\
                                           &                                                             &                                                             &                                                             &                                                             &                                                             &                                                             &                                                             &                                                             \\
\multirow{-2}{*}{Italy}                    & \multirow{-2}{*}{-}                                         & \multirow{-2}{*}{-}                                         & \multirow{-2}{*}{-}                                         & \multirow{-2}{*}{-}                                         & \multirow{-2}{*}{-}                                         & \multirow{-2}{*}{-}                                         & \multirow{-2}{*}{-}                                         & \multirow{-2}{*}{-}                                         \\
                                           &                                                             &                                                             &                                                             &                                                             &                                                             &                                                             & \cellcolor[HTML]{C6E0B4}                                    & \cellcolor[HTML]{C6E0B4}                                    \\
\multirow{-2}{*}{Latvia}                   & \multirow{-2}{*}{-}                                         & \multirow{-2}{*}{-}                                         & \multirow{-2}{*}{-}                                         & \multirow{-2}{*}{-}                                         & \multirow{-2}{*}{-}                                         & \multirow{-2}{*}{-}                                         & \multirow{-2}{*}{\cellcolor[HTML]{C6E0B4}\textgreater 90\%} & \multirow{-2}{*}{\cellcolor[HTML]{C6E0B4}\textgreater 90\%} \\
                                           &                                                             &                                                             & \cellcolor[HTML]{C6E0B4}                                    & \cellcolor[HTML]{C6E0B4}                                    & \cellcolor[HTML]{C6E0B4}                                    & \cellcolor[HTML]{C6E0B4}                                    & \cellcolor[HTML]{C6E0B4}                                    & \cellcolor[HTML]{C6E0B4}                                    \\
\multirow{-2}{*}{Lithuania}                & \multirow{-2}{*}{-}                                         & \multirow{-2}{*}{-}                                         & \multirow{-2}{*}{\cellcolor[HTML]{C6E0B4}\textgreater 90\%} & \multirow{-2}{*}{\cellcolor[HTML]{C6E0B4}\textgreater 90\%} & \multirow{-2}{*}{\cellcolor[HTML]{C6E0B4}\textgreater 90\%} & \multirow{-2}{*}{\cellcolor[HTML]{C6E0B4}\textgreater 90\%} & \multirow{-2}{*}{\cellcolor[HTML]{C6E0B4}\textgreater 90\%} & \multirow{-2}{*}{\cellcolor[HTML]{C6E0B4}\textgreater 90\%} \\
                                           &                                                             &                                                             & \cellcolor[HTML]{C6E0B4}                                    & \cellcolor[HTML]{C6E0B4}                                    & \cellcolor[HTML]{C6E0B4}                                    & \cellcolor[HTML]{C6E0B4}                                    & \cellcolor[HTML]{C6E0B4}                                    & \cellcolor[HTML]{C6E0B4}                                    \\
\multirow{-2}{*}{Luxemburg}                & \multirow{-2}{*}{-}                                         & \multirow{-2}{*}{-}                                         & \multirow{-2}{*}{\cellcolor[HTML]{C6E0B4}\textgreater 90\%} & \multirow{-2}{*}{\cellcolor[HTML]{C6E0B4}\textgreater 90\%} & \multirow{-2}{*}{\cellcolor[HTML]{C6E0B4}\textgreater 90\%} & \multirow{-2}{*}{\cellcolor[HTML]{C6E0B4}\textgreater 90\%} & \multirow{-2}{*}{\cellcolor[HTML]{C6E0B4}\textgreater 90\%} & \multirow{-2}{*}{\cellcolor[HTML]{C6E0B4}\textgreater 90\%} \\
                                           &                                                             &                                                             &                                                             &                                                             &                                                             & \cellcolor[HTML]{C6E0B4}                                    & \cellcolor[HTML]{C6E0B4}                                    & \cellcolor[HTML]{C6E0B4}                                    \\
\multirow{-2}{*}{Macedonia}                & \multirow{-2}{*}{-}                                         & \multirow{-2}{*}{-}                                         & \multirow{-2}{*}{-}                                         & \multirow{-2}{*}{-}                                         & \multirow{-2}{*}{-}                                         & \multirow{-2}{*}{\cellcolor[HTML]{C6E0B4}\textgreater 90\%} & \multirow{-2}{*}{\cellcolor[HTML]{C6E0B4}\textgreater 90\%} & \multirow{-2}{*}{\cellcolor[HTML]{C6E0B4}\textgreater 90\%} \\
                                           &                                                             &                                                             &                                                             &                                                             &                                                             &                                                             &                                                             &                                                             \\
\multirow{-2}{*}{Netherlands}              & \multirow{-2}{*}{-}                                         & \multirow{-2}{*}{-}                                         & \multirow{-2}{*}{-}                                         & \multirow{-2}{*}{-}                                         & \multirow{-2}{*}{-}                                         & \multirow{-2}{*}{-}                                         & \multirow{-2}{*}{-}                                         & \multirow{-2}{*}{-}                                         \\
                                           & \cellcolor[HTML]{C6E0B4}                                    & \cellcolor[HTML]{C6E0B4}                                    & \cellcolor[HTML]{C6E0B4}                                    & \cellcolor[HTML]{C6E0B4}                                    & \cellcolor[HTML]{C6E0B4}                                    & \cellcolor[HTML]{C6E0B4}                                    & \cellcolor[HTML]{C6E0B4}                                    & \cellcolor[HTML]{C6E0B4}                                    \\
\multirow{-2}{*}{Norway}                   & \multirow{-2}{*}{\cellcolor[HTML]{C6E0B4}\textgreater 90\%} & \multirow{-2}{*}{\cellcolor[HTML]{C6E0B4}\textgreater 90\%} & \multirow{-2}{*}{\cellcolor[HTML]{C6E0B4}\textgreater 90\%} & \multirow{-2}{*}{\cellcolor[HTML]{C6E0B4}\textgreater 90\%} & \multirow{-2}{*}{\cellcolor[HTML]{C6E0B4}\textgreater 90\%} & \multirow{-2}{*}{\cellcolor[HTML]{C6E0B4}\textgreater 90\%} & \multirow{-2}{*}{\cellcolor[HTML]{C6E0B4}\textgreater 90\%} & \multirow{-2}{*}{\cellcolor[HTML]{C6E0B4}\textgreater 90\%} \\
                                           &                                                             &                                                             &                                                             &                                                             &                                                             &                                                             &                                                             &                                                             \\
\multirow{-2}{*}{Poland}                   & \multirow{-2}{*}{-}                                         & \multirow{-2}{*}{-}                                         & \multirow{-2}{*}{-}                                         & \multirow{-2}{*}{-}                                         & \multirow{-2}{*}{-}                                         & \multirow{-2}{*}{-}                                         & \multirow{-2}{*}{-}                                         & \multirow{-2}{*}{-}                                         \\
                                           &                                                             &                                                             &                                                             &                                                             &                                                             & \cellcolor[HTML]{C6E0B4}                                    & \cellcolor[HTML]{C6E0B4}                                    & \cellcolor[HTML]{C6E0B4}                                    \\
\multirow{-2}{*}{Portugal}                 & \multirow{-2}{*}{-}                                         & \multirow{-2}{*}{-}                                         & \multirow{-2}{*}{-}                                         & \multirow{-2}{*}{-}                                         & \multirow{-2}{*}{-}                                         & \multirow{-2}{*}{\cellcolor[HTML]{C6E0B4}\textgreater 90\%} & \multirow{-2}{*}{\cellcolor[HTML]{C6E0B4}\textgreater 90\%} & \multirow{-2}{*}{\cellcolor[HTML]{C6E0B4}\textgreater 90\%} \\
                                           &                                                             &                                                             &                                                             &                                                             &                                                             &                                                             & \cellcolor[HTML]{C6E0B4}                                    & \cellcolor[HTML]{C6E0B4}                                    \\
\multirow{-2}{*}{Romania}                  & \multirow{-2}{*}{-}                                         & \multirow{-2}{*}{-}                                         & \multirow{-2}{*}{-}                                         & \multirow{-2}{*}{-}                                         & \multirow{-2}{*}{-}                                         & \multirow{-2}{*}{-}                                         & \multirow{-2}{*}{\cellcolor[HTML]{C6E0B4}\textgreater 90\%} & \multirow{-2}{*}{\cellcolor[HTML]{C6E0B4}\textgreater 90\%} \\
                                           &                                                             &                                                             &                                                             &                                                             &                                                             &                                                             &                                                             & \cellcolor[HTML]{C6E0B4}                                    \\
\multirow{-2}{*}{Serbia}                   & \multirow{-2}{*}{-}                                         & \multirow{-2}{*}{-}                                         & \multirow{-2}{*}{-}                                         & \multirow{-2}{*}{-}                                         & \multirow{-2}{*}{-}                                         & \multirow{-2}{*}{-}                                         & \multirow{-2}{*}{-}                                         & \multirow{-2}{*}{\cellcolor[HTML]{C6E0B4}\textgreater 90\%} \\
                                           &                                                             &                                                             &                                                             &                                                             &                                                             &                                                             &                                                             &                                                             \\
\multirow{-2}{*}{Slovakia}                 & \multirow{-2}{*}{-}                                         & \multirow{-2}{*}{-}                                         & \multirow{-2}{*}{-}                                         & \multirow{-2}{*}{-}                                         & \multirow{-2}{*}{-}                                         & \multirow{-2}{*}{-}                                         & \multirow{-2}{*}{-}                                         & \multirow{-2}{*}{-}                                         \\
                                           &                                                             &                                                             &                                                             &                                                             &                                                             &                                                             &                                                             &                                                             \\
\multirow{-2}{*}{Slovenia}                 & \multirow{-2}{*}{-}                                         & \multirow{-2}{*}{-}                                         & \multirow{-2}{*}{-}                                         & \multirow{-2}{*}{-}                                         & \multirow{-2}{*}{-}                                         & \multirow{-2}{*}{-}                                         & \multirow{-2}{*}{-}                                         & \multirow{-2}{*}{-}                                         \\
                                           &                                                             &                                                             &                                                             &                                                             &                                                             & \cellcolor[HTML]{C6E0B4}                                    & \cellcolor[HTML]{C6E0B4}                                    & \cellcolor[HTML]{C6E0B4}                                    \\
\multirow{-2}{*}{Spain}                    & \multirow{-2}{*}{-}                                         & \multirow{-2}{*}{-}                                         & \multirow{-2}{*}{-}                                         & \multirow{-2}{*}{-}                                         & \multirow{-2}{*}{-}                                         & \multirow{-2}{*}{\cellcolor[HTML]{C6E0B4}\textgreater 90\%} & \multirow{-2}{*}{\cellcolor[HTML]{C6E0B4}\textgreater 90\%} & \multirow{-2}{*}{\cellcolor[HTML]{C6E0B4}\textgreater 90\%} \\
                                           &                                                             &                                                             &                                                             &                                                             & \cellcolor[HTML]{C6E0B4}                                    & \cellcolor[HTML]{C6E0B4}                                    & \cellcolor[HTML]{C6E0B4}                                    & \cellcolor[HTML]{C6E0B4}                                    \\
\multirow{-2}{*}{Sweden}                   & \multirow{-2}{*}{-}                                         & \multirow{-2}{*}{-}                                         & \multirow{-2}{*}{-}                                         & \multirow{-2}{*}{-}                                         & \multirow{-2}{*}{\cellcolor[HTML]{C6E0B4}\textgreater 90\%} & \multirow{-2}{*}{\cellcolor[HTML]{C6E0B4}\textgreater 90\%} & \multirow{-2}{*}{\cellcolor[HTML]{C6E0B4}\textgreater 90\%} & \multirow{-2}{*}{\cellcolor[HTML]{C6E0B4}\textgreater 90\%} \\
                                           &                                                             &                                                             &                                                             &                                                             &                                                             &                                                             &                                                             &                                                             \\
\multirow{-2}{*}{Switzerland}              & \multirow{-2}{*}{-}                                         & \multirow{-2}{*}{-}                                         & \multirow{-2}{*}{-}                                         & \multirow{-2}{*}{-}                                         & \multirow{-2}{*}{-}                                         & \multirow{-2}{*}{-}                                         & \multirow{-2}{*}{-}                                         & \multirow{-2}{*}{-}                                        
\end{tabular}
\end{table}

\end{document}